\documentclass[10pt,twocolumn,twoside]{IEEEtran}
\usepackage{graphicx}
\usepackage{subfigure}
\usepackage{color}
\usepackage{multicol}
\usepackage{amssymb}
\usepackage{multirow}
\usepackage{amsmath}
\usepackage{bm}
\usepackage{cite}
\usepackage{gensymb}
\usepackage{amsfonts}
\usepackage{mathrsfs}
\usepackage{amsmath}
\usepackage{algorithm}
\usepackage{algorithmic}
\usepackage{amsthm}

\usepackage{tabularx}
\usepackage{balance}
\usepackage{mathrsfs}
\usepackage{hyperref}
\usepackage{array}
\usepackage{multirow}
\usepackage{CJK}
\newcommand{\PreserveBackslash}[1]{\let\temp=\\#1\let\\=\temp}
\newcolumntype{C}[1]{>{\PreserveBackslash\centering}p{#1}}
\newcolumntype{L}[1]{>{\PreserveBackslash\raggedright}p{#1}}
\newcolumntype{R}[1]{>{\PreserveBackslash\raggedleft}p{#1}}
\usepackage[flushleft]{threeparttable}

\begin{document}
	
\bibliographystyle{IEEEtran} 
	
\title{Wideband Beam Tracking in THz Massive MIMO Systems}
	
\author{
		Jingbo Tan, \emph{Student Member, IEEE}  and Linglong Dai, \emph{Senior Member, IEEE}
		
		\thanks{All authors are with the Beijing National Research Center for Information Science and Technology (BNRist)	as well as the Department of Electronic Engineering, Tsinghua University, Beijing 100084, P. R. China (E-mails: tanjb17@mails.tsinghua.edu.cn; daill@tsinghua.edu.cn).}
		\thanks{This work was supported in part by the National Key Research and Development Program of China (Grant No. 2020YFB1807201), in part by the National Natural Science Foundation of China (Grant No. 62031019), and in part by the European Commission through the H2020-MSCA-ITN META WIRELESS Research Project under Grant 956256.}
		\vspace{-4mm}
	}
	
\maketitle
\IEEEpeerreviewmaketitle

\begin{abstract}
	Terahertz (THz) massive multiple-input multiple-output (MIMO) has been considered as one of the promising technologies for future 6G wireless communications. It is essential to obtain channel information by beam tracking scheme to track mobile users in THz massive MIMO systems. However, the existing beam tracking schemes designed for narrowband systems with the traditional hybrid precoding structure suffer from a severe performance loss caused by the beam split effect, and thus cannot be directly applied to wideband THz massive MIMO systems. To solve this problem, in this paper we propose a beam zooming based beam tracking scheme by considering the recently proposed delay-phase precoding structure for THz massive MIMO. Specifically, we firstly prove the beam zooming mechanism to flexibly control the angular coverage of frequency-dependent beams over the whole bandwidth, i.e., the degree of the beam split effect, which can be realized by the elaborate design of time delays in the delay-phase precoding structure. Then, based on this beam zooming mechanism, we propose to track multiple user physical directions simultaneously in each time slot by generating multiple beams. The angular coverage of these beams is flexibly zoomed to adapt to the potential variation range of the user physical direction. After several time slots, the base station is able to obtain the exact user physical direction by finding out the beam with the largest user received power. Unlike traditional schemes where only one frequency-independent beam can be usually generated by one radio-frequency chain, the proposed beam zooming based beam tracking scheme can simultaneously track multiple user physical directions by using multiple frequency-dependent beams generated by one radio-frequency chain. Theoretical analysis shows that the proposed scheme can achieve the near-optimal achievable sum-rate performance with low beam training overhead, which is also verified by extensive simulation results.
\end{abstract}
	
\begin{IEEEkeywords}
	THz massive MIMO, beam tracking, hybrid precoding.\vspace{-3mm}
\end{IEEEkeywords}

\section{Introduction}
Terahertz (THz) communication is considered as one of the promising technologies for future 6G wireless communications, since it can provide tens of GHz bandwidth to support ultra-high data rates\cite{Ref:6GUser2020,Ref:Survey6G2019,Ref:THzChannel2019,Ref:THzTimingAc2017,Ref:Survey6G2018}. However, THz signals suffer from the severe path loss due to the high carrier frequencies\cite{Ref:THzWaveDe2016}. To compensate for the severe path loss, massive multiple-input multiple-output (MIMO), which can generate directional beams with high array gains, is considered promising to be integrated in future THz communications\cite{Ref:TeraSub2016,Ref:AoATera2018,Ref:BDMATera2017}. Nevertheless, the widely considered hybrid precoding structure in massive MIMO \cite{Ref:SpatiallyPre2014} cannot deal with the beam split effect caused by the wide bandwidth and a large number of antennas in THz massive MIMO systems\cite{Ref:WideBeamCE2019}. Specifically, the beam split effect can be seen as a serious situation of the widely known beam squint\cite{Ref:EffWide2018,Ref:WideCode2019}, which means that the beams generated by the traditional \emph{frequency-independent} phase-shifters (PSs) may be totally split into different physical directions over different subcarriers within the large frequency band. Consequently, these beams over different subcarriers cannot be aligned with the target user in a certain direction, which leads to a serious array gain loss and thus an obvious achievable sum-rate loss. To solve this problem, introducing time-delayers into precoding structure, such as true-time-delay array\cite{Ref:TTD_Filter2017,Ref:TTDarray2019,Ref:TTDtrack2020}, array-of-subarray structure\cite{Ref:SubarrayCode2017}, and delay-phase precoding structure\cite{Ref:DPP2019}, is considered to be promising. Thanks to the \emph{frequency-dependent} phase shifts provided by time-delayers, these precoding structures can significantly mitigate the array gain loss caused by the beam split effect.
	
To realize precoding, accurate channel information is essential. Generally, the channel information can be obtained through channel estimation. However, because of the large size of channel information, traditional channel estimation schemes will result in an unacceptable channel estimation overhead in THz massive MIMO systems\cite{Ref:DisCSchannelES2014}. To avoid such an unacceptable overhead, the beam training scheme is preferred. Instead of estimating full channel information of large size, the beam training scheme directly estimates the physical directions of channel paths\cite{Ref:BeamTra2020}, which is realized by using directional beams through a training procedure between the base station (BS) and users. Thanks to the quasi-optical characteristic of THz channel\cite{Ref:ThzCh2007} and the accurate physical directions obtained by beam training, the beam selection based precoding method is able to achieve the near-optimal achievable sum-rate when users are quasi-static\cite{Ref:BeamSe2018,Ref:BeamDSeML2019,Ref:NObeamSe2016}. Unfortunately, the beam training scheme suffers from a high training overhead when users are moving. Specifically, since the optimal beam of a moving user varies fast due to the narrow beam width, the beam training procedure has to be carried out frequently, and thus results in a high beam training overhead. Therefore, to reduce the beam training overhead for mobile users, an efficient beam tracking scheme is required for practical THz massive MIMO systems\cite{Ref:RoBeaMTra2017,Ref:BeamTra2017,Ref:TrackTera2017,Ref:BeamAlignment2010,Ref:Mill2013,Ref:BeamPairT2018}.
	
\subsection{Prior Works}
The existing beam tracking schemes can be generally divided into two categories. The first category mainly relies on the user mobility model\cite{Ref:TrackTera2017,Ref:BeamTra2017,Ref:RoBeaMTra2017}. The second category is codebook-based beam tracking, where a training procedure between the BS and the user is carried out to find out the optimal beam from a predefined beam codebook\cite{Ref:BeamAlignment2010,Ref:Mill2013,Ref:BeamPairT2018}.
	
For the first category of beam tracking schemes, the key problem is how to model the user mobility. Specifically, \cite{Ref:RoBeaMTra2017} assumed that the user mobility satisfies the first-order Gauss-Markov model, and an extended Kalman filter method was proposed to track the optimal beam. To improve the beam tracking accuracy, the user mobility was further formulated as a kinematic model, and a modified unscented Kalman filter was exploited to track the channel angles more accurately\cite{Ref:BeamTra2017}. In addition, based on the linear motion model defined by user physical direction and user velocity, a priori-aided beam tracking scheme was proposed in \cite{Ref:TrackTera2017}. Nevertheless, this category of beam tracking schemes highly relies on the user mobility model as a priori, which maybe inaccurate and cannot be easily obtained, especially in THz massive MIMO systems. 
	
The second category of beam tracking schemes depends on the design of codebook-based beam training algorithms, where each codeword in the codebook determines a directional beam. For instance, \cite{Ref:BeamAlignment2010} searched the optimal beam among a beam codebook containing potential beams through a single-sided exhausted training procedure. To reduce the unacceptable beam training overhead caused by the large codebook size in \cite{Ref:BeamAlignment2010}, an adaptive search scheme was proposed by using the hierarchical codebook, which consists of different beam codewords with different angular coverages \cite{Ref:Mill2013}. To further accelerate the beam tracking procedure, an auxiliary beam pair based beam tracking scheme was proposed in \cite{Ref:BeamPairT2018}, where the optimal beam was obtained based on the user received signals of two auxiliary beams generated by two extra RF chains. Note that codebook-based beam tracking schemes have been widely considered in millimeter-wave massive MIMO systems \cite{Ref:BeamCodebook2018}.

Although the existing beam tracking schemes above \cite{Ref:RoBeaMTra2017,Ref:BeamTra2017,Ref:TrackTera2017,Ref:BeamAlignment2010,Ref:Mill2013,Ref:BeamPairT2018} can achieve the acceptable performance, they are only suitable for \emph{narrowband} systems with the traditional hybrid precoding structure. In \emph{wideband} THz massive MIMO systems, since the hybrid precoding structure cannot mitigate the serious beam split effect, these schemes will suffer from a severe performance degradation. Consequently, an efficient wideband beam tracking scheme is essential for wideband THz massive MIMO systems. Recently, several wideband beam tracking or training scheme have been proposed. Specifically, a fast tracking scheme based on frequency-dependent beams generated by true-time-delay array was proposed in \cite{Ref:TTDtrack2020}. While, due to the large number of antennas, utilizing true-time-delay array will introduce unacceptable energy consumption in THz massive MIMO. \cite{Ref:SubarrayCode2017} proposed a codebook based beam training scheme based on an array-of-subarray structure with a reduced number of time-delayers. Nevertheless, the codebook in \cite{Ref:SubarrayCode2017} is fixed and cannot be well adapted to user motion. In addition, \cite{Ref:THzShotNat2020} proposed a fast wideband tracking scheme in THz communications by utilizing the property that wideband THz signals emitted from a single leaky waveguide will split into different directions. However, the scheme in \cite{Ref:THzShotNat2020} can only be adopted when a single leaky waveguide is employed, which will cause a limitation on transmission distance. Therefore, to the best of our knowledge, the wideband beam tracking problem for wideband THz massive MIMO has not been addressed in the literature.
	
\subsection{Our Contributions}
To fill in this gap, we propose a beam zooming based beam tracking scheme to solve the wideband beam tracking problem in THz massive MIMO systems. For the wideband systems, the severe performance loss caused by the beam split effect must be eliminated. Thus, in this paper we consider the delay-phase precoding structure\cite{Ref:DPP2019}, which has been proved to be able to achieve the near-optimal achievable sum-rate performance with acceptable energy consumption in wideband THz massive MIMO systems. The contributions of this paper can be summarized as follows.
\begin{itemize}
	\item We reveal the beam zooming mechanism by analyzing the angular coverage of \emph{frequency-dependent} beams generated by the delay-phase precoding structure. We show that by the elaborate design of time delays, i.e., the frequency-dependent phase shifts, the angular coverage of these beams can be flexibly zoomed to achieve a required angular range. This mechanism to flexibly control the angular coverage, i.e., the degree of beam split effect, enables us to generate multiple beams simultaneously by using only one RF chain, which is impossible for existing schemes.
	\item Based on the beam zooming mechanism, we propose a beam zooming based beam tracking scheme to solve the wideband beam tracking problem. In the proposed scheme, multiple user physical directions are tracked by multiple frequency-dependent beams in each time slot. By leveraging the beam zooming mechanism, the angular coverage of these beams can be flexibly controlled to cover a fraction of the potential variation range of the user physical direction. After the whole variation range of the user physical direction has been tracked, the BS can obtain the optimal beam based on the user received signal power. Unlike traditional beam tracking schemes which usually track only one user physical direction in each time slot, the proposed scheme is able to track multiple user physical directions in each time slot by actively controlling the angular coverage of frequency-dependent beams, i.e, the degree of beam split effect. Thus, the beam training overhead can be significantly reduced. 
	\item We further provide the theoretical analysis of the required beam training overhead and achievable sum-rate performance of the proposed scheme. The relationship between the achievable sum-rate and the beam tracking accuracy will also be revealed. The analysis shows that, the proposed scheme can achieve the near-optimal achievable sum-rate with very low beam training overhead, which is supported by extensive simulation results\footnote{Simulation codes are provided to reproduce the results presented in this paper: http://oa.ee.tsinghua.edu.cn/dailinglong/publications/publications.html.}.
\end{itemize}
	
\subsection{Organization and Notation}
The remainder of this paper is organized as follows. In Section \ref{Sys}, the system model of wideband THz massive MIMO systems is introduced. In Section \ref{TraBeamTra}, we first discuss the direct application of the typical beam tracking scheme \cite{Ref:BeamAlignment2010} by using the delay-phase precoding structure. Then, the beam zooming mechanism to flexibly control the angular coverage is revealed, based on which we propose a beam zooming based beam tracking scheme to realize efficient beam tracking with low beam training overhead. In Section \ref{Per}, theoretical analysis of beam training overhead and achievable sum-rate is provided. Section \ref{Sim} shows the simulation results. Finally, conclusions are drawn in Section \ref{Con}.
	
\emph{Notation:} Lower-case and upper-case boldface letters represent vectors and matrices, respectively; $(\cdot)^{T}$, $(\cdot)^{H}$, $\|\cdot\|_{F}$, and $\|\cdot\|_{k}$ denote the transpose, conjugate transpose, Frobenius norm, and $k$-norm of a matrix, respectively; $\mathbf{H}_{[i,j]}$ denotes the element of matrix $\mathbf{H}$ at the $i$-th row and the $j$-th column; $\mathbb{E}(\cdot)$ denotes the expectation; $|\cdot|$ denotes the absolute operator; $\mathbf{I}_{N}$ represents the identity matrix of size $N\times N$; $\mathrm{blkdiag}(\mathbf{A})$ denotes a block diagonal matrix, where columns of $\mathbf{A}$ represent the diagonal blocks of the matrix $\mathrm{blkdiag}(\mathbf{A})$ in order; $\mathcal{CN}(\mathbf{\mu},\mathbf{\Sigma})$ denotes the Gaussian distribution with mean $\mathbf{\mu}$ and covariance $\mathbf{\Sigma}$; Finally, $\mathcal{U}(a,b)$ denotes the uniform distribution between $a$ and $b$.
	
\section{System Model}\label{Sys}
In this section, we introduce the system model of the wideband THz massive MIMO system. Specifically, the wideband THz ray-based channel model is illustrated at first. Then, the delay-phase precoding structure and the corresponding precoding design is explained. Finally, we present the widely considered frame structure for beam tracking.
	
\subsection{Channel Model}
In this paper, a multi-user wideband THz massive MIMO system is considered. The BS with an $N$-antenna uniform linear array is employed to serve $K$ single-antenna users by using orthogonal frequency division multiplexing (OFDM) with $M$ subcarriers\footnote{We consider the uniform linear array for simplifying the expression in this paper. Please note that our proposed scheme can be also utilized in THz massive MIMO systems with uniform planar array.}. The bandwidth is denoted as $B$. We consider the widely used ray-based channel model for wideband THz channel \cite{Ref:TeraSub2016}. Specifically, the downlink channel of the $k$-th user at the $m$-th subcarrier $\mathbf{h}_{k,m}\in\mathcal{C}^{1\times N}$ with $k=1,2,\cdots,K$ and $m=1,2,\cdots,M$ can be denoted as
\begin{equation}\label{1}
\mathbf{h}_{k,m}=\beta^{(0)}_{k,m}\mathbf{a}_{N}^{H}\left(\psi_{k,m}^{(0)}\right)+\sum_{l=1}^{L-1}\beta^{(l)}_{k,m}\mathbf{a}_{N}^{H}\left(\psi_{k,m}^{(l)}\right),
\end{equation}
where $\beta_{k,m}^{(l)}=g_{k,m}^{(l)}e^{-j\pi\tau_{k}^{(l)}f_{m}}$ for $l=0,1,\cdots,L-1$ with $g_{k,m}^{(l)}$ and $\tau_{k}^{(l)}$ being the path gain and the time delay of the $l$-th path for user $k$ respectively,  $f_{m}$ denotes the frequency of the $m$-th subcarrier satisfying $f_{m}=f_{c}+\frac{B}{M}(m-1-\frac{M-1}{2})$ with $f_{c}$ being the central frequency, $L$ denotes the number of paths, $\psi_{k,m}^{(0)}$ is the spatial direction of the line-of-sight (LoS) path of the $k$-th user at subcarrier $m$, $\psi_{k,m}^{(l)}$ with $l=1,2,\cdots,L-1$ is the spatial directions of the non-LoS (NLoS) paths of the $k$-th user at subcarrier $m$,  and $\mathbf{a}_{N}(\psi_{k,m}^{(l)})\in\mathcal{C}^{N\times 1}$ denotes the array response vector presented as 
\begin{equation}
\mathbf{a}_{N}(\psi_{k,m}^{(l)})=\frac{1}{\sqrt{N}}[1,e^{j\pi\psi_{k,m}^{(l)}},e^{j\pi 2\psi_{k,m}^{(l)}},\cdots,e^{j\pi(N-1)\psi_{k,m}^{(l)}}]^{T}, 
\end{equation}
where the \emph{spatial directions} $\psi_{k,m}^{(l)}$ satisfy $\psi_{k,m}^{(l)}=\frac{2d}{c}f_{m}\sin{\tilde{\theta}_{k}^{(l)}}$ for $l=0,1,\cdots,L$ with $\tilde{\theta}_{k}^{(l)}$ being the \emph{physical direction} of the $l$-th path for the $k$-th user, $d$ is the antenna spacing usually set as half of the wavelength of the central frequency, i.e., $d=\lambda_{c}/2=f_{c}/2c$ with $\lambda_{c}$ denoting the wavelength at the central frequency and $c$ being the speed of light. For simplification, we use $\theta_{k}^{(l)}=\sin{\tilde{\theta}_{k}^{(l)}}$ to represent the user physical direction in this paper, whose value range is $\theta_{k}^{(l)}\in[-1,1]$.

In (\ref{1}), we have considered the frequency-dependent path gain $g_{k,m}^{(l)}$. Here, we focus on the path gain of the LoS path $g_{k,m}^{(0)}$. The free space path loss (FSPL) model can be utilized to model the path gain of the LoS path \cite{Ref:THzChannel2019}. Specifically, the path gain of the LoS path at frequency $f$ and transmission distance $D$ satisfies the following model \cite{Ref:THzChannel2019}  as
\begin{equation}\label{model1}
\begin{aligned}
&\left(g_{k,m}^{(0)}\right)^{2}[\mathrm{dB}]\\
&=\mathrm{FSPL}(f,D)[\mathrm{dB}]=32.4+20\log_{10}(f)+20\log_{10}D. 
\end{aligned}
\end{equation}
Considering that transmission distance $D$ is the same at different frequencies for an arbitrary user, we can obtain the relationship between the path gain at subcarrier $m$ and central frequency $f_\mathrm{c}$ as
\begin{equation}\label{model2}
g_{k,m}^{(0)}=\frac{f_{m}}{f_\mathrm{c}}g_{k,\mathrm{c}}^{(0)},
\end{equation}
where $g_{k,\mathrm{c}}^{(0)}$ is the path gain at the central frequency which is assumed to be a real number without loss of generality.

\begin{figure}
	\centering
	\includegraphics[width=0.46\textwidth]{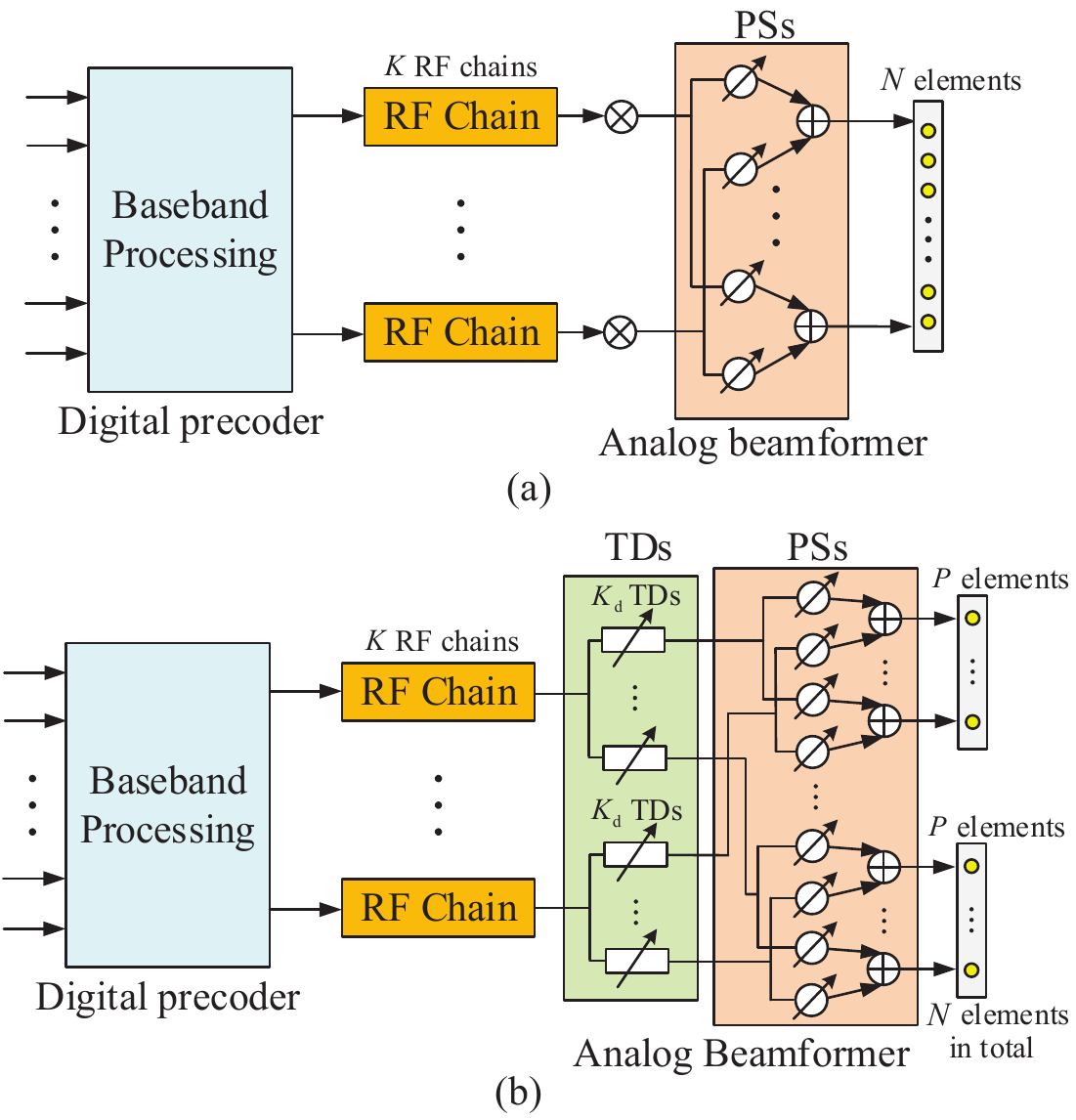}
	\caption{Two different precoding structures: (a)  Hybrid precoding structure; (b) Delay-phase precoding structure\cite{Ref:DPP2019}.}
	\vspace{-2mm}
\end{figure}
\begin{figure}
	\centering
	{\includegraphics[width=0.47\textwidth]{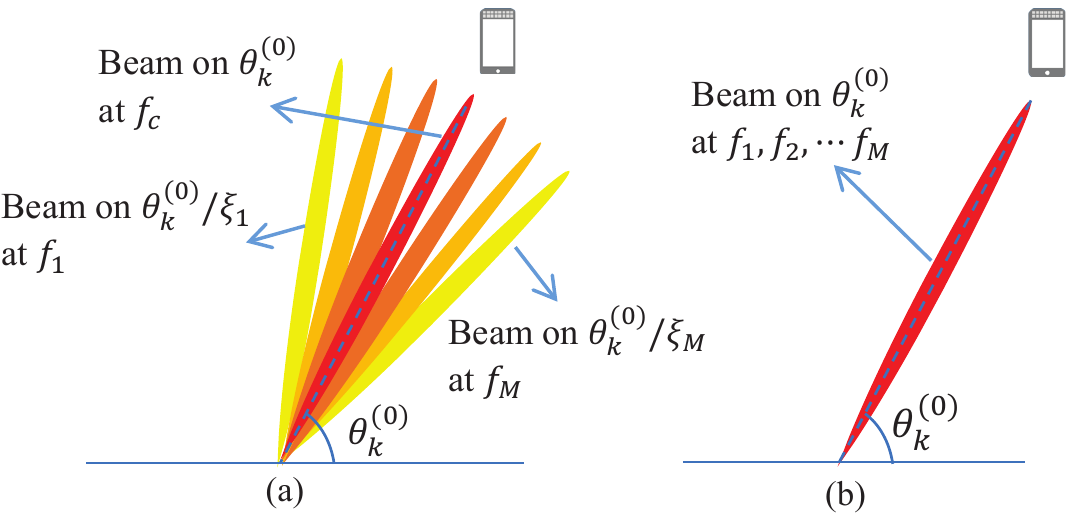}}
	\centering{\caption{Beam split effect and beams generated by the DPP structure: (a) The beam split effect; (b) Beams generated by the DPP structure.}}
	\vspace{-3mm}
\end{figure}
	
Hybrid precoding structure as shown in Fig 1. (a) is widely considered in THz massive MIMO systems, since it can generate high-array-gain directional beams with acceptable energy consumption\cite{Ref:SpatiallyPre2014}. Due to the severe loss incurred by the scattering, THz communication heavily relies on the LoS path\cite{Ref:ThzCh2007}, so the beam selection based  precoding method is near-optimal for multi-user THz massive MIMO systems\cite{Ref:BeamSe2018,Ref:BeamDSeML2019,Ref:NObeamSe2016}, where each user is served through a beam aligned with its physical direction $\theta_{k}^{(0)}$ of the LoS path. Generally, the directional beam for each user is generated by \emph{frequency-independent} phase-shifters (PSs) in hybrid precoding structure. To be more specific, the beamforming vector $\mathbf{f}_{k}$ for the $k$-th user is usually set as $\mathbf{f}_{k}=\mathbf{a}_{N}(\theta_{k}^{(0)})$\cite{Ref:NObeamSe2016} to generate a directional beam aligned with the physical direction of the LoS path of the $k$-th user. 
	
However, as shown in Fig. 2 (a), the beam generated by the \emph{frequency-independent} beamforming vector $\mathbf{f}_{k}=\mathbf{a}_{N}(\theta_{k}^{(0)})$ in wideband THz massive MIMO may split into different physical directions at different subcarrier frequencies, which is caused by the ultra-wide bandwidth and large antenna number. This effect is called as beam split effect \cite{Ref:DPP2019}\footnote{The mechanism of the beam split and the beam squint \cite{Ref:EffWide2018,Ref:WideCode2019} is similar. While, the beam split can be seen as a serious situation of the beam squint which may be more easier to occur in THz massive MIMO systems as illustrated in \cite{Ref:DPP2019}.} . Specifically, as proved by \textbf{Lemma 1} in \cite{Ref:DPP2019}, the beam generated by the \emph{frequency-independent} beamforming vector $\mathbf{f}_{k}=\mathbf{a}_{N}(\theta_{k}^{(0)})$ will be aligned with the frequency-dependent physical direction $\theta_{k,m}$ at subcarrier $m$ as
\begin{equation}\label{3}
\theta_{k,m}=\left(f_{c}/f_{m}\right)\theta_{k}^{(0)}=\theta_{k}^{(0)}/\xi_{m},
\end{equation}
where we define $\xi_{m}=f_{m}/f_{c}$ as the relative frequency compared with the central frequency $f_{c}$. It is clear from (\ref{3}) that the beam generated by the beamforming vector $\mathbf{f}_{k}$ may point to frequency-dependent physical directions at different subcarrier frequencies $f_{m}$. Considering the ultra-wide bandwidth and the very narrow beam generated by large number of antennas in THz massive MIMO systems, the frequency-independent beam generated by $\mathbf{f}_{k}=\mathbf{a}_{N}(\theta_{k}^{(0)})$ cannot be aligned with the user physical direction $\theta_{k}^{(0)}$ at most of the subcarriers, as shown in Fig. 2 (a). As a result, the traditional hybrid precoding structure will suffer from a severe achievable sum-rate loss within the ultra-wide bandwidth, e.g., it is shown that more than $60$\% performance loss will be introduced when the bandwidth is increased from $1$ GHz in mmWave systems to $10$ GHz in THz systems\cite{Ref:DPP2019}. 
	
\subsection{Delay-Phase Precoding}\label{DPP}
To cope with the severe performance loss caused by the beam split effect, we have recently proposed the delay-phase precoding (DPP) structure in \cite{Ref:DPP2019}. In the DPP structure, a time-delay network is introduced as a new precoding layer between RF chains and PSs network as shown in Fig. 1 (b), which transforms the \emph{frequency-independent} analog beamforming into the \emph{frequency-dependent} analog beamforming by utilizing the frequency-dependent phase shifts provided by time delays. It has been proved that the DPP structure is able to mitigate the severe achievable sum-rate loss caused by the beam split effect with an acceptable energy consumption\cite{Ref:DPP2019}.
	
Therefore, in this paper, we consider the DPP sturture at the BS, where each RF chain connects to all the antenna elements via time-delayers (TDs). Specifically, each RF chain connects to $K_\mathrm{d}$ TDs, and these $K_\mathrm{d}$ TDs will be connected to all $N$ antenna elements via PSs without overlapping, i.e., the $j$-th TD is connected to $P$ antenna elements with index from $(j-1)P+1$ to $jP$ via $P$ PSs, where $P=N/K_\mathrm{d}$ is assumed to be an integer. In this paper, we also set the number of RF chains $N_\mathrm{RF}$ equal to the number of users $K$, i.e., $N_\mathrm{RF}=K$\cite{Ref:NObeamSe2016}. Then, the received signal $\mathbf{y}_{m}\in\mathcal{C}^{K\times 1}$ at the $m$-th subcarrier for all $K$ single-antenna users can be denoted as
\begin{equation}\label{4}
\mathbf{y}_{m}=\mathbf{H}_{m}\mathbf{A}_{m}\mathbf{D}_{m}\mathbf{s}+\mathbf{n},
\end{equation}
where $\mathbf{H}_{m}=[\mathbf{h}_{1,m}^{T},\mathbf{h}_{2,m}^{T},\cdots,\mathbf{h}_{K,m}^{T}]^{T}\in\mathcal{C}^{K\times N}$ denotes the downlink channel matrix between the BS and $K$ users at the $m$-th subcarrier, $\mathbf{A}_{m}\in\mathcal{C}^{N\times K}$ denotes the frequency-dependent analog beamformer realized by PSs and TDs, $\mathbf{D}_{m}\in\mathcal{C}^{K\times K}$ is the digital precoder satisfying the power constraint $\|\mathbf{A}_{m}\mathbf{D}_{m,[:,k]}\|_{F}\leq\rho$ with $\rho$ being the transmission power for each user, and $\mathbf{n}\in\mathcal{C}^{K\times1}$ is the AWGN noise following the distribution $\mathcal{CN}(0,\sigma^{2}\mathbf{I}_{K})$ with $\sigma^{2}$ presenting the noise power. By introducing TDs, the analog beamformer $\mathbf{A}_{m}$ is separated into two parts as $\mathbf{A}_{m}=\mathbf{A}^\mathrm{s}\mathbf{A}_{m}^\mathrm{d}$. On one hand, $\mathbf{A}^\mathrm{s}\in\mathcal{C}^{N\times K_\mathrm{d}K}=[\mathbf{A}_{1}^\mathrm{s},\mathbf{A}_{2}^\mathrm{s},\cdots,\mathbf{A}_{K}^\mathrm{s}]$ is realized by frequency-independent PSs, where $\mathbf{A}_{k}^\mathrm{s}\in\mathcal{C}^{N\times K_\mathrm{d}}=\mathrm{blkdiag}([\mathbf{a}_{k,1},\mathbf{a}_{k,2},\cdots,\mathbf{a}_{k,K}])$ with $\mathbf{a}_{i,j},i,j=1,2,\cdots,K$ denoting the beamforming vector provided by the $P$ PSs connected to the $j$-th TD corresponding to the $i$-th RF chain. Note that due to the use of PSs, each element of $\mathbf{a}_{i,j}$ should satisfy constant amplitude constraint as  $|\mathbf{a}_{i,j,[p,q]}|=\frac{1}{\sqrt{N}}$\cite{Ref:OverMilliMIMO2016}. On the other hand,  $\mathbf{A}_{m}^\mathrm{d}\in\mathcal{C}^{K_\mathrm{d}K\times K}$ is realized by TDs, which can be represented as
\begin{equation}\label{5}
\mathbf{A}_{m}^\mathrm{d}=\mathrm{blkdiag}\left([e^{-j2\pi f_{m}\mathbf{t}_{1}},e^{-j2\pi f_{m}\mathbf{t}_{2}},\cdots,e^{-j2\pi f_{m}\mathbf{t}_{K}}]\right),
\end{equation}
where $\mathbf{t}_{i}\in\mathcal{C}^{K_\mathrm{d}\times1},i=1,2,\cdots,K$ contains the time delays provided by the $K_\mathrm{d}$ TDs that connected to the $i$-th RF chain. From (\ref{5}), we know that the TDs introduce frequency-dependent phase shifts into the analog beamformer $\mathbf{A}_{m}$, which enables the frequency-dependent analog beamforming. Without loss of generality, we define the $k$-th column of the analog beamformer $\mathbf{A}_{m}$ as the frequency-dependent beamforming vector for the $k$-th user at subcarrier $m$, which can be denoted as $\mathbf{f}_{k,m}=\mathbf{A}_{[:,k]}=\mathbf{A}^\mathrm{s}_{k}e^{-j2\pi f_{m}\mathbf{t}_{k}}$. The criterion on how to choose the optimal value of the number of TDs $K_\mathrm{d}$ is important for the design of the DPP structure, which has been well adressed in the previous work \cite{Ref:DPP2019}.
	
As illustrated in Fig. 2 (b), by utilizing the frequency-dependent phase shifts in (\ref{5}), the DPP structure can mitigate the beam split effect by generating beams aligned with the user over the whole bandwidth. To realize this goal, \cite{Ref:DPP2019} has proved the principle to design the frequency-dependent beamforming vectors $\mathbf{f}_{k,m}$, i.e., the PSs still generate a beam aligned with the user physical direction $\theta_{k}^{(0)}$, while time delays provided by TDs are elaborately designed to rotate beams over different subcarrier frequencies to the user physical direction $\theta_{k}^{(0)}$. Specifically,  to mitigate the beam split effect for the $k$-th user, the analog beamformer $\mathbf{A}^\mathrm{s}_{k}$ which contains phase shifts provided by PSs and the time delays provided by TDs $\mathbf{t}_{k}$ should follow the form as \cite{Ref:DPP2019}
\begin{equation}\label{6}
\mathbf{A}_{k}^\mathrm{s}=\mathrm{blkdiag}\Big([\underbrace{\mathbf{a}_{P}(\theta_{k}^{(0)}),\mathbf{a}_{P}(\theta_{k}^{(0)}),\cdots,\mathbf{a}_{P}(\theta_{k}^{(0)})}_{K_\mathrm{d}~\text{columns}}]\Big),
\end{equation}
\begin{equation}\label{7}
\mathbf{t}_{k}=s_{k}T_{c}\mathbf{p}(K_\mathrm{d}),
\end{equation}
\begin{equation}\label{8}
s_{k}=-\left(P\theta_{k}^{(0)}\right)/2,
\end{equation}
where $T_{c}=\frac{1}{f_{c}}$ is the period at the central frequency $f_{c}$, $s_{k}$ denotes the number of periods that delayed by TDs, and $\mathbf{p}\left(K_\mathrm{d}\right)$ is defined as $\mathbf{p}\left(K_\mathrm{d}\right)=[0,1,\cdots,K_\mathrm{d}-1]^{T}$. Based on (\ref{6}), (\ref{7}) and (\ref{8}), the frequency-dependent beamforming vector $\mathbf{f}_{k,m}=\mathbf{A}^\mathrm{s}_{k}e^{-j2\pi f_{m}\mathbf{t}_{k}}$ can generate beams aligned with the user physical direction over the whole bandwidth. Thus, the severe performance loss caused by the beam split effect can be eliminated, which makes the DPP structure promising for wideband THz massive MIMO systems. 
	
\subsection{Frame Structure}
In this subsection, we introduce the frame structure as shown in Fig. 3, which has been widely used in the literature for THz communications\cite{Ref:Mill2013,Ref:TraAoD2016}. Two factors that influence the THz channel have been considered: 1) the disappearance or appearance of the LoS path due to environmental changes, e.g., blockage by the vehicles; 2) the change of the user physical direction caused by the user mobility.
	
\begin{figure}
	\centering
	{\includegraphics[width=0.49\textwidth]{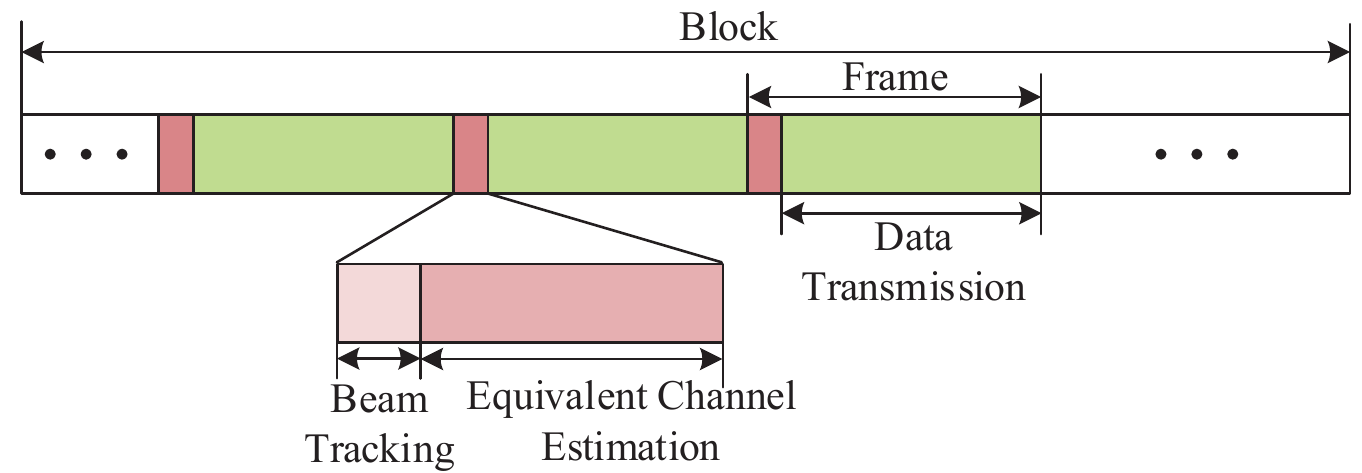}}
	\centering{\caption{The frame structure for THz communications.}}
	\vspace{-3mm}
\end{figure}
	
Firstly, the time period between the successive LoS path disappearance or appearance is defined as a block, and the estimation of the full channel for each user is carried out at the beginning of each block. Then, in a block, the variation of the physical direction of the LoS path induces the change of the optimal beam in a much smaller time scale, which is defined as a frame. At the beginning of each frame, the beam tracking scheme is carried out, where new user physical directions are tracked through a beam training procedure. This beam training procedure usually lasts several time slots. In different time slots, the BS transmits training pilot sequences to users by using different beamforming vectors. Based on new user physical directions obtained by beam tracking, the analog beamformer $\mathbf{A}_{m}$ can be determined to generate directional beams for each user according to (\ref{6}), (\ref{7}) and (\ref{8}). Thus, the received signal $\mathbf{y}_{m}$ at users can be converted to 
\begin{equation}
\mathbf{y}_{m}=\mathbf{H}_{m,\mathrm{eq}}\mathbf{D}_{m}\mathbf{s}+\mathbf{n},
\end{equation}
where $\mathbf{H}_{m,\mathrm{eq}}=\mathbf{H}_{m}\mathbf{A}_{m}$ is the equivalent channel. Since the equivalent channel $\mathbf{H}_{m,\mathrm{eq}}$ of size $K\times K$ is low-dimensional, it can be estimated through traditional channel estimation methods with a low overhead during the equivalent channel estimation period. After that, the digital precoder $\mathbf{D}_{m}$ can be determined based on the equivalent channel $\mathbf{H}_{m,\mathrm{eq}}$ by using existing precoding methods, e.g., zero-forcing (ZF)\cite{Ref:FundWC2005} or minimum mean square error (MMSE) \cite{Ref:MMSEPre2019}. Finally, the user data is transmitted by using the analog beamformer $\mathbf{A}_{m}$ and the digital precoder $\mathbf{D}_{m}$ during the data transmission period.
	
In the above frame structure, the accuracy of beam tracking scheme has a crucial impact on the achievable sum-rate performance. Unfortunately, the existing beam tracking schemes\cite{Ref:RoBeaMTra2017,Ref:BeamTra2017,Ref:TrackTera2017,Ref:BeamAlignment2010,Ref:Mill2013,Ref:BeamPairT2018} designed for narrowband systems with the traditional hybrid precoding structure cannot deal with the beam split effect, which will result in a serious achievable sum-rate performance loss in wideband THz massive MIMO systems. Therefore, an efficient wideband beam tracking scheme is required for wideband THz massive MIMO systems.
	
\section{The Proposed Flexible Beam Zooming Based Beam Tracking Method}\label{TraBeamTra}
In this section, to solve the wideband beam tracking problem in wideband THz massive MIMO systems, we propose a beam zooming based beam tracking scheme by exploiting the DPP structure. Specifically, we first discuss the direct application of a typical beam tracking scheme\cite{Ref:BeamAlignment2010} using the DPP structure, which serves as a benchmark for comparison. Then, we prove the beam zooming mechanism that different user physical directions can be tracked simultaneously by flexibly controlling the degree of the beam split effect. Finally, based on the beam zooming mechanism, a beam zooming based beam tracking scheme is proposed with low beam training overhead.

\subsection{Typical Beam Tracking Scheme Using DPP}\label{Tra}
\begin{figure}
	\centering
	\includegraphics[width=0.47\textwidth]{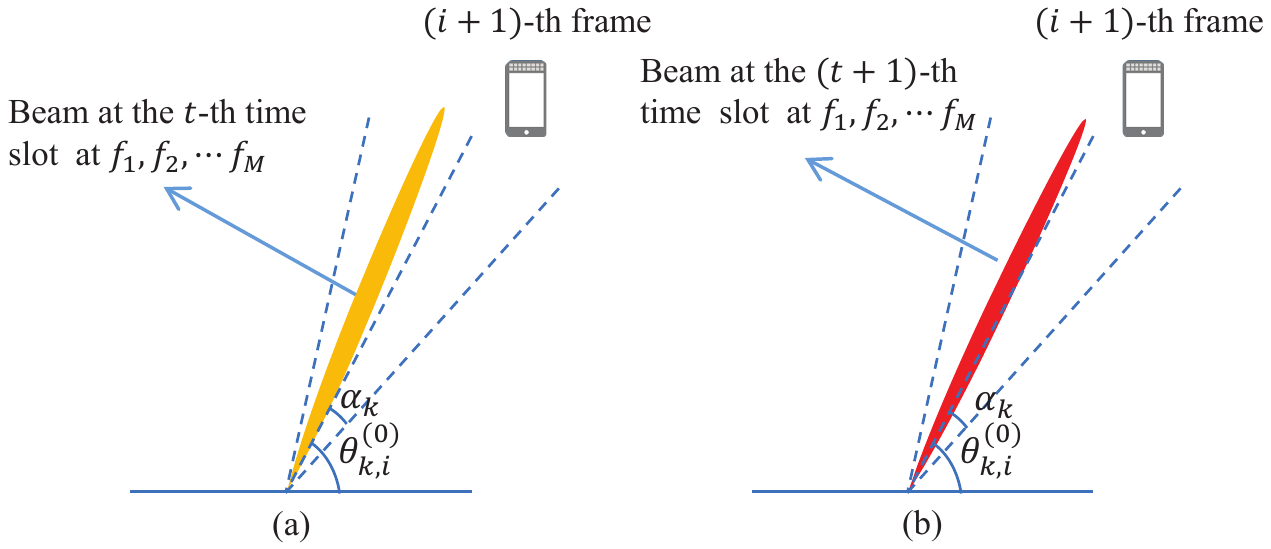}
	\caption{The typical beam tracking scheme \cite{Ref:BeamAlignment2010} adapted to the DPP structure: (a) the beams generated by the DPP structure at the $t$-th time slot; (b) the beams generated by the DPP structure at the $(t+1)$-th time slot;.\vspace{-3mm}}
\end{figure}
	
In the typical beam tracking scheme \cite{Ref:BeamAlignment2010}, the optimal beamforming vector for a specific user is selected out from a beam codebook through a training procedure between the BS and the user. Each codeword in the beam codebook determines a potential beam aligned with a unique physical direction. In different time slots during beam tracking, the BS transmits training pilot sequence to each user using different codewords in the codebook. Then, the codeword with the largest received power is selected out as the beamforming vector for the next frame. However, the typical beam tracking scheme \cite{Ref:BeamAlignment2010} is designed for narrowband systems with the traditional hybrid precoding structure, and it cannot be directly applied in the DPP structure. Therefore, we first discuss the adaptation of the typical beam tracking scheme\cite{Ref:BeamAlignment2010} to the DPP structure, which can serve as the benchmark for comparison.
	
The physical direction of the $k$-th user at the $i$-th frame can be denoted as $\theta_{k,i}^{(0)}$, and the target of beam tracking is to track the physical direction at the $(i+1)$-th frame $\theta_{k,i+1}^{(0)}$ based on $\theta_{k,i}^{(0)}$. Generally, the prior information of user mobility can be exploited to narrow the angular tracking range. In this paper,  we assume a simple prior information that the BS knows the potential variation range of user physical direction $\alpha_{k}$ for the $k$-th user, which means $\theta_{k,i+1}^{(0)}$ lies in an angular tracking range as $[\theta_{k,i}^{(0)}-\alpha_{k},\theta_{k,i}^{(0)}+\alpha_{k}]$. This variation range $\alpha_{k}$ can be obtained through control signaling or efficient user trajectory prediction in advance\cite{Ref:TraPreML2019}. 
	
Based on the above description, the procedure of the typical beam tracking scheme \cite{Ref:BeamAlignment2010} using the DPP structure can be explained as follows. We denote $T$ as the beam training overhead, i.e., the number of time slots used for beam tracking. In the $t$-th time slot with $t=1,2,\cdots,T$, the beamforming vector $\mathbf{f}_{k,m}$ is designed according to (\ref{6}), (\ref{7}) and (\ref{8}) to form a beam aligned with the physical direction $\bar{\theta}_{k,i}^{(t)}$ over the whole bandwidth for the $k$-th user, as shown in Fig. 4. The physical direction $\bar{\theta}_{k,i}^{(t)}$ in each time slot $t$ satisfies
\begin{equation}\label{10.1}
\bar{\theta}_{k,i}^{(t)}=\theta_{k,i}^{(0)}-\alpha_{k}+(2t-1)\frac{\alpha_{k}}{T}.
\end{equation}
In this way, these physical directions $\bar{\theta}_{k,i}^{(t)},t=1,2,\cdots,T$ can uniformly cover the angular tracking range $[\theta_{k,i}^{(0)}-\alpha_{k},\theta_{k,i}^{(0)}+\alpha_{k}]$. Then, the BS transmits pilot sequence using the beam aligned with physical direction $\bar{\theta}_{k,i}^{(t)}$ to the $k$-th user in the $t$-th time slot. After $T$ time slots, the physical direction $\theta_{k,i+1}^{(0)}$ can be selected out from $T$ physical directions $\bar{\theta}_{k,i}^{(t)},t=1,2,\cdots,T$ by finding the physical direction corresponding to the largest user received power. Finally, a beam aligned with $\theta_{k,i+1}^{(0)}$ can be generated to serve the user in the $(i+1)$-th frame. Note that in the DPP structure, $K$ beams for all $K$ users can be generated at the same time as described in Subsection \ref{DPP}. Therefore, the beam tracking procedure for all $K$ users can be carried out simultaneously.
	
The typical beam tracking scheme using the DPP structure can realize beam tracking without the loss of the achievable sum-rate performance. Nevertheless, since the typical beam tracking scheme searches the tracking range exhaustively, the required beam training overhead is too large to achieve sufficient beam tracking accuracy, especially when the narrow beam width is considered in THz massive MIMO systems. For example, when $N=256$ and $\alpha_{k}=0.1$, the beam training overhead $T$ should be larger than $52$, which is unacceptable in practice. Consequently, the solution to realize low-overhead beam tracking is essential for THz massive MIMO systems.

\begin{figure}
	\centering
	\includegraphics[width=0.49\textwidth]{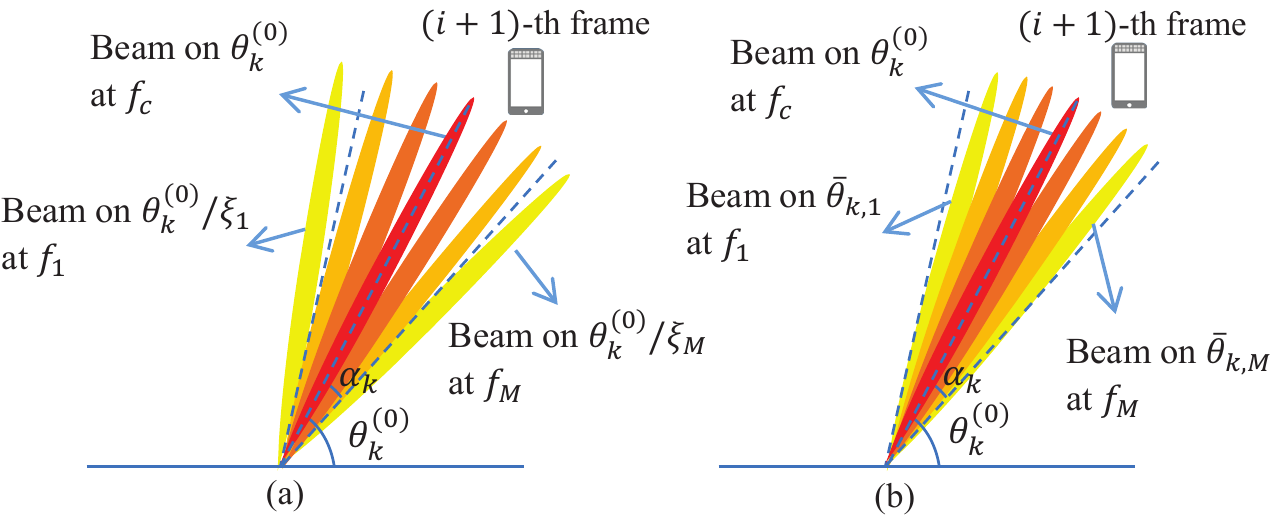}
	\caption{Beam zooming mechanism: (a) The beams without control by TDs; (b) The beams generated based on \textbf{Lemma 2}.}\vspace{-3mm}
\end{figure}

\vspace{-2mm}
\subsection{Beam Zooming Mechanism}\label{FBZ}
The main reason for the high beam training overhead of the typical beam tracking scheme \cite{Ref:BeamAlignment2010} is that, only one physical direction can be tracked in each time slot. This cannot be easily solved by the traditional hybrid precoding structure, since only one frequency-independent beam can be generated by using frequency-independent PSs connected to one RF chain over the whole bandwidth. In contrast, by introducing time delays, the DPP structure is able to not only mitigate the beam split effect, but also to flexibly control the angular coverage of the generated frequency-dependent beams, i.e, control the degree of the beam split effect. If the beams at different subcarriers are aligned with different user physical directions to cover the angular tracking range $[\theta_{k,i}^{(0)}-\alpha_{k},\theta_{k,i}^{(0)}+\alpha_{k}]$, multiple potential user physical directions can be tracked simultaneously. In this way, a much lower beam training overhead can be expected. Inspired by the above idea, in this subsection we reveal the beam zooming mechanism that can flexibly control the angular coverage of the frequency-dependent beams generated by the DPP structure as shown in Fig. 5. 
	
To better reveal the beam zooming mechanism, we first provide a useful \textbf{Lemma 1} as the basis of the beam zooming mechanism, where we ignore the frame index $i$ for simplification. \textbf{Lemma 1} describes how the DPP structure adjusts the physical direction that the frequency-dependent beam is aligned with. Specifically, the physical direction that the generated beam is aligned with can be flexibly adjusted from the physical direction determined by the beam split effect to a new physical direction, which is determined by the frequency-dependent phase shifts provided by TDs.
	
\newtheorem{thm}{Lemma}
\begin{thm} \label{lemma2}
	When the analog beamformer $\mathbf{A}_{k}^\mathrm{s}$ which contains phase shifts provided by PSs is designed to generate a beam aligned with physical direction $\phi_{k}$ as $\mathbf{A}_{k}^\mathrm{s}=\mathrm{blkdiag}\left(\mathbf{a}_{P}(\phi_{k})e^{j\pi P\phi_{k}\mathbf{p}^{T}(K_\mathrm{d})}\right)$ and time delays provided by TDs satisfies $e^{-j2\pi f_{m}\mathbf{t}_{k}}=\left[1,e^{j\pi\beta_{k,m}},e^{j\pi2\beta_{k,m}},\cdots,e^{j\pi(K_{d}-1)\beta_{k,m}}\right]^{T}$, the beamforming vector $\mathbf{f}_{k,m}=\mathbf{A}_{k}^\mathrm{s}e^{-j2\pi f_{m}\mathbf{t}_{k}}$ at the $m$-th subcarrier frequency $f_{m}$ can form a beam aligned with the physical direction $\bar{\theta}_{k,m}$ as
	\begin{equation}\label{App1}
	\bar{\theta}_{k,m}=\arg\max\limits_{\theta}\eta(\mathbf{f}_{k,m},\theta,f_{m})=\frac{\phi_{k}}{\xi_{m}}+\frac{\beta_{k,m}}{\xi_{m}P},
	\end{equation}
	where $\beta_{k,m}\in[-1,1]$ denotes the frequency-dependent phase shifts provided by TDs, $P=N/ K_\mathrm{d}$, and $\eta(\mathbf{f}_{k,m},\theta,f_{m})$ denotes the array gain achieved by the beamforming vector $\mathbf{f}_{k,m}$ on an arbitrary physical direction $\theta\in[-1,1]$ at the $m$-th subcarrier frequency $f_{m}$. Moreover, the array gain achieved by $\mathbf{f}_{k,m}$ satisfies
	\begin{equation}\label{App2}
	\begin{aligned}
	\eta(\mathbf{f}_{k,m},\theta,f_{m})=\frac{1}{N}|\Xi_{K_d}&(P(\phi_{k}-\xi_{m}\theta)+\beta_{k,m})\\
	&\times\Xi_{P}(\phi_{k}-\xi_{m}\theta)|,
	\end{aligned}
	\end{equation}
	where $\Xi_{\bar{N}}(\alpha)=\sin\frac{\bar{N}\pi}{2}\alpha/\bar{N}\sin\frac{\pi}{2}\alpha$ is the Dirchlet sinc function.
\end{thm}
\emph{Proof}: By transforming the physical direction $\phi_{k}$ to its spatial direction $\frac{\phi_{k}}{\xi_{m}}$ at the $m$-th subcarrier, (\ref{App1}) and (\ref{App2}) can be easily proved by using \textbf{Lemma 2} in \cite{Ref:DPP2019}.$\hfill\blacksquare$
	
\textbf{Lemma 1} reveals that, the physical direction $	\bar{\theta}_{k,m}$ that the beam generated by beamforming vector $\mathbf{f}_{k,m}$ is aligned with, can be flexibly adjusted by setting different frequency-dependent phase shifts $\beta_{k,m}$ provided by TDs. It is worth noting that the range of the physical direction adjustment decided by the frequency-dependent phase shift $\beta_{k,m}$ (i.e., the second term $\frac{\beta_{k,m}}{\xi_{m}P}$ in (\ref{App1})) is limited by the value range of $\beta_{k,m}$ as $\beta_{k,m}\in[-1,1]$. When $\beta_{k,m}\in[-1,1]$ is not satisfied, the beamforming vector $\mathbf{f}_{k,m}$ cannot generate a directional beam with sufficient array gain, which has been proved by using (\ref{App2}) in \cite{Ref:DPP2019}.
	
Based on \textbf{Lemma 1}, we further reveal the beam zooming mechanism in the following \textbf{Lemma 2}. It proves that by the elaborate design of phase shifts provided by PSs and TDs, the angular coverage of beams generated by frequency-dependent beamforming vectors $\mathbf{f}_{k,m},m=1,2,\cdots,M$ can be flexibly zoomed.

\begin{thm} \label{lemma1}
	Consider the $k$-th user and define a physical direction $\phi_{k}$ as $\phi_{k}=\theta_{k}^{(0)}+(1-\xi_{1})\alpha_{k}$. When the number of periods that delayed by TDs is $s_{k}=-\frac{P}{2}\left(\phi_{k}+\frac{2\xi_{M}\xi_{1}\alpha_{k}}{\xi_{M}-\xi_{1}}\right)$, the analog beamformer satisfies $\mathbf{A}_{k}^\mathrm{s}=\mathrm{blkdiag}\left(\mathbf{a}_{P}(\phi_{k})e^{j\pi (P\phi_{k}+2s_{k})\mathbf{p}^{T}(K_\mathrm{d})}\right)$ and the time delays provided by TDs  $\mathbf{t}_{k}=s_{k}T_{c}\mathbf{p}(K_\mathrm{d})$, the beamforming vector  $\mathbf{f}_{k,m}=\mathbf{A}_{k}^{s}e^{-j2\pi f_{m}\mathbf{t}_{k}}$ for the $k$-th user at the $m$-th subcarrier can form a beam aligned with the physical direction $\bar{\theta}_{k,m}$ as
	\begin{equation}\label{10}
	\bar{\theta}_{k,m}=\theta_{k}^{(0)}+(1-\xi_{1})\alpha_{k}+\frac{2\xi_{M}\xi_{1}(\xi_{m}-1)}{\xi_{m}(\xi_{M}-\xi_{1})}\alpha_{k}.
	\end{equation}
	Moreover, the $M$ physical directions $\bar{\theta}_{k,m}, m=1,2,\cdots,M$ can cover the whole angular tracking range $[\theta_{k}^{(0)}-\alpha_{k},\theta_{k}^{(0)}+\alpha_{k}]$ of the $k$-th user .
\end{thm}
\emph{Proof:} See Appendix A. $\hfill\blacksquare$

\begin{figure}
	\centering
		\includegraphics[width=0.45\textwidth]{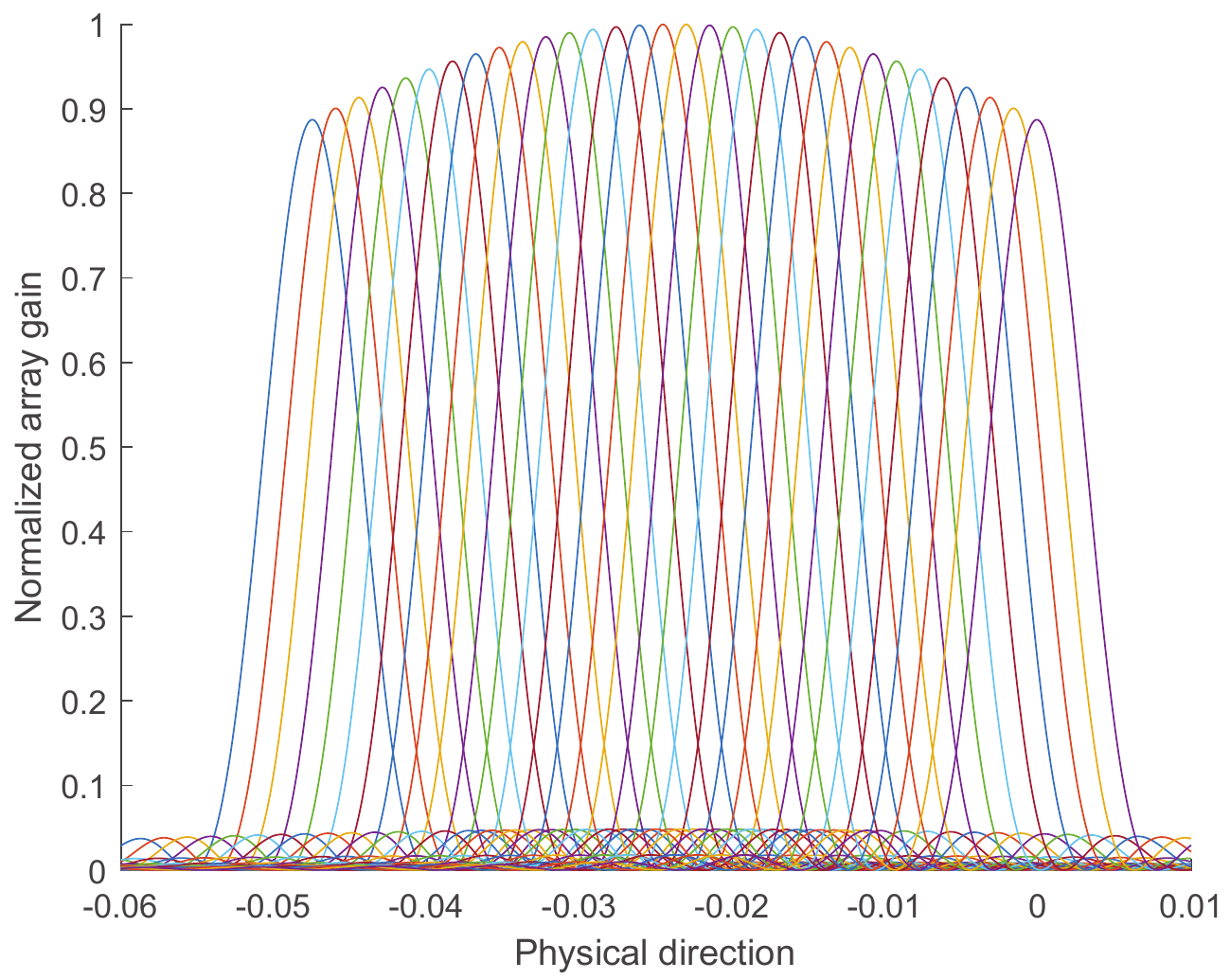}
		\caption{The beam pattern of beams generated by the beam zooming mechanism with parameters $\theta_{k}^{(0)}=-0.025$, $\alpha_{k}=0.025$, $M=32$, $N=256$, $f_\mathrm{c}=100$ GHz, $B=10$ GHz.}\vspace{-2mm}
\end{figure}
	
\begin{figure}
	\centering
	\includegraphics[width=0.44\textwidth]{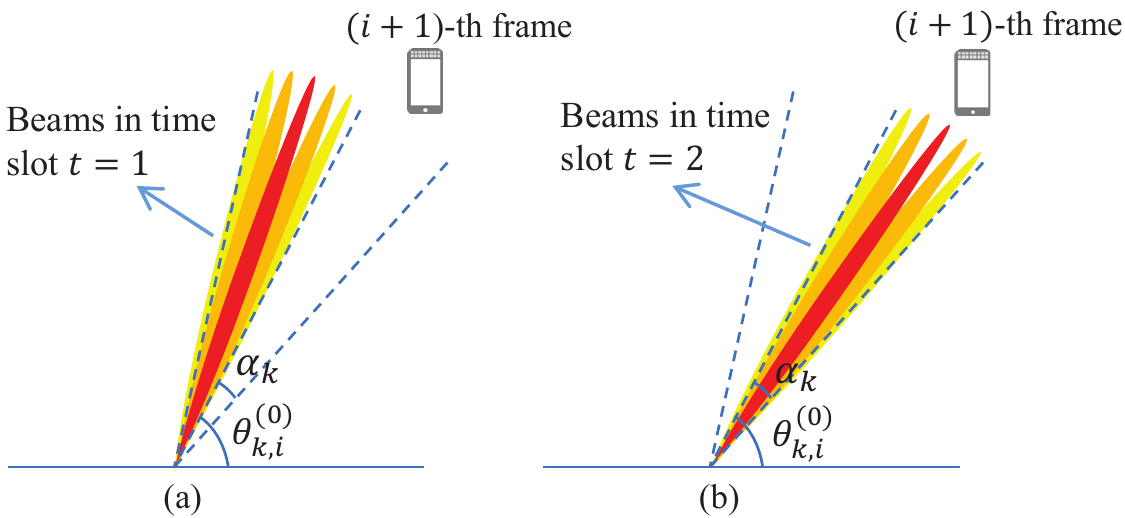}
	\caption{ The proposed beam zooming based beam tracking scheme with training overhead $T=2$: (a) Beams generated in time slot $t=1$; (b)  Beams generated in time slot $t=2$.}\vspace{-2mm}
\end{figure}
	
The beam zooming mechanism proved in \textbf{Lemma 2} illustrates that through the elaborate design on phase shifts and time delays, the angular coverage of beams can be flexibly controlled according to the required angular tracking range as shown in Fig. 5 (b), but not determined by the beam split effect which is shown in Fig. 5 (a). To give a better illustration of the beam zooming mechanism, we provide the beam pattern of generated beams at different subcarriers in Fig. 6, with parameters $\theta_{k}^{(0)}=-0.025$, $\alpha_{k}=0.025$, $M=32$, $N=256$, $f_\mathrm{c}=100$ GHz, and $B=10$ GHz. We can observe from Fig. 6 that the beam zooming mechanism generates correct beams with angle-domain coverage $[-0.05,0]$. Therefore, we can conclude that the beam zooming mechanism can be utilized to generate frequency-dependent beams to track different user physical directions at the same time.

\subsection{Beam Zooming Based Beam Tracking Scheme}\label{FBZAF}
	
Based on the beam zooming mechanism above, we propose a beam zooming based beam tracking scheme in this subsection. The key idea of the beam zooming based scheme is shown in Fig. 7. In each time slot, the angular coverage of frequency-dependent beams generated by the DPP structure is flexibly zoomed to adapt to the angular tracking range based on the beam zooming mechanism. Specifically, these beams are designed to cover a fraction of the required angular tracking range. Then, multiple user physical directions in this fraction of angular tracking range are tracked simultaneously. The above procedure is carried out for $T$ times in all $T$ time slots until the whole angular tracking range is tracked. Finally, the physical direction of each user in the next frame can be obtained by finding out the physical direction corresponding to the largest user received signal power.
	
\begin{algorithm}[htb]
	\caption{Proposed beam zooming based beam tracking scheme.}
	\label{alg:Framwork}
	\begin{algorithmic}[1]
		\REQUIRE ~~\\
		Physical directions $\theta_{k,i}^{(0)}$; Variation range of user physical direction $\alpha_{k}$; Beam tracking overhead $T$; The number of pilots in each time slot $Q$; The number of TDs connected to a RF chain $K_{d}$;\\
		\ENSURE ~~\\
		Physical directions $\theta_{k,i+1}^{(0)}$\\
		\STATE $\bar{\theta}_{k,i,\mathrm{cen}}^{(t)}=\theta_{k,i}^{(0)}-\alpha_{k}+\frac{(2t-1)\alpha_{k}}{T}$\\
		\STATE $\bar{\theta}_{k,m,i}^{(t)}=\bar{\theta}_{k,i,\mathrm{cen}}^{(t)}+(1-\xi_{1})\frac{\alpha_{k}}{T}+\frac{2\xi_{M}\xi_{1}(\xi_{m}-1)}{\xi_{m}(\xi_{M}-\xi_{1})}\frac{\alpha_{k}}{T}$\\
		\STATE $\mathbf{\Psi}_{k}^{i+1}=[\bar{\theta}_{k,1,i}^{(1)},\bar{\theta}_{k,2,i}^{(1)},\cdots,\bar{\theta}_{k,M,i}^{(1)},\bar{\theta}_{k,1,i}^{(2)},\bar{\theta}_{k,2,i}^{(2)}\cdots,\bar{\theta}_{k,M,i}^{(T)}]$\\
		\FOR{$t\in\{1,2,\cdots,T\}$}		
		\STATE 
		$\phi_{k}^{(t)}=\bar{\theta}_{k,i,\mathrm{cen}}^{(t)}+(1-\xi_{1})\frac{\alpha_{k}}{T}$\\
		\STATE $s_{k}^{(t)}=-\frac{P}{2}\left(\phi_{k}^{(t)}+\frac{2\xi_{M}\xi_{1}\alpha_{k}}{(\xi_{M}-\xi_{1})T}\right)$\\
		\STATE $\mathbf{A}_{k}^{\mathrm{s},(t)}=\mathrm{blkdiag}\left(\mathbf{a}_{P}(\phi_{k}^{(t)})e^{j\pi (P\phi_{k}^{(t)}+2s_{k}^{(t)})\mathbf{p}^{T}(K_\mathrm{d})}\right)$\\
		\STATE $\mathbf{t}_{k}=s_{k}^{(t)}T_{c}\mathbf{p}(K_\mathrm{d})$\\
		\STATE
		$\mathbf{f}_{k,m}^{(t)}=\mathbf{A}_{k}^{\mathrm{s},(t)}e^{-j2\pi f_{m}\mathbf{t}_{k}^{(t)}}$\\
		\STATE $\mathbf{A}_{m}^{(t)}=\left[\mathbf{f}_{1,m}^{(t)},\mathbf{f}_{2,m}^{(t)},\cdots,\mathbf{f}_{K,m}^{(t)}\right]$\\
		\STATE
		 $\mathbf{Y}_{m,t}=\kappa_{m}\mathbf{H}_{m}\mathbf{A}^{(t)}_{m}\mathbf{Q}^{(t)}_{m}+\mathbf{N}^{(t)}$
		\ENDFOR 
		\STATE $\left(t_{k},m_{k}\right)=\mathop{\arg\max}\limits_{t\in{1,2,\cdots,T},m\in{1,2,\cdots,M}}\|\mathbf{Y}_{m,t,[k,:]}\mathbf{q}_{k,m}^{(t)}\|_{2}^{2}$\\
		\STATE $\theta_{k,i+1}^{(0)}=\mathbf{\Psi}^{t+1}_{k,[(t_{k}-1)M+m_{k}]}$
		\RETURN $\theta_{k,i+1}^{(0)}$.
	\end{algorithmic}
\end{algorithm}

The specific procedure of the proposed beam zooming based beam tracking scheme is provided in \textbf{Algorithm 1}. The target of \textbf{Algorithm 1} is to obtain the physical directions of users $\theta_{k,i+1}^{(0)}$ at the $(i+1)$-th frame based on the user physical directions at the $i$-th frame $\theta_{k,i}^{(0)}$ for all $K$ users $k=1,2,\cdots,K$. At first, the user physical directions that will be tracked in the $t$-th time slot are calculated in steps $1$ and $2$, where $\bar{\theta}_{k,i,\mathrm{cen}}^{(t)}$ denotes the central physical direction of the $t$-th fraction of the angular tracking range $[\theta_{k,i}^{(0)}-\alpha_{k}+\frac{(2t-2)\alpha_{k}}{T},\theta_{k,i}^{(0)}-\alpha_{k}+\frac{2t\alpha_{k}}{T}]$, and $\bar{\theta}_{k,m,i}^{(t)}$ denotes the physical direction that will be tracked at the $m$-th subcarrier in the $t$-th time slot as
\begin{equation}\label{15}
\bar{\theta}_{k,m,i}^{(t)}=\bar{\theta}_{k,i,\mathrm{cen}}^{(t)}+(1-\xi_{1})\frac{\alpha_{k}}{T}+\frac{2\xi_{M}\xi_{1}(\xi_{m}-1)}{\xi_{m}(\xi_{M}-\xi_{1}){T}}\alpha_{k}.
\end{equation}
Then, all potential user physical directions that will be tracked in $T$ time slots are combined as a target physical direction set $\mathbf{\Psi}_{k}^{i+1}$ in step $3$.

After the target physical direction set has been built, these physical directions are tracked in $T$ time slots to guarantee the whole angular tracking range will be searched. In the $t$-th time slot, beams that cover the $t$-th fraction of angular tracking range $[\theta_{k,i}^{(0)}-\alpha_{k}+\frac{(2t-2)\alpha_{k}}{T},\theta_{k,i}^{(0)}-\alpha_{k}+\frac{2t\alpha_{k}}{T}]$ are generated by the DPP structure in steps $5-10$, where the analog beamformer $\mathbf{A}_{m}$ is calculated based on the parameters in \textbf{Lemma 2}. Specifically, in steps $5$ and $6$, $\phi_{k}^{(t)}$ and $s_{k}^{(t)}$ is calculated as
\begin{equation}\label{16}
\phi_{k}^{(t)}=\bar{\theta}_{k,i,\mathrm{cen}}^{(t)}+(1-\xi_{1})\frac{\alpha_{k}}{T},
\end{equation}
\begin{equation}\label{17}
s_{k}^{(t)}=-\frac{P}{2}\left(\phi_{k}^{(t)}+\frac{2\xi_{M}\xi_{1}\alpha_{k}}{(\xi_{M}-\xi_{1})T}\right).
\end{equation}
Then, in steps $7$ and $8$, phase shifts provided by frequency-independent PSs $\mathbf{A}_{k}^{\mathrm{s},(t)}$ and time delays provided by TDs $\mathbf{t}_{k}^{(t)}$ are designed. As proved in \textbf{Lemma 2}, when $\phi_{k}^{(t)}$ and $s_{k}^{(t)}$ satisfy (\ref{16}) and (\ref{17}), the beams generated by $\mathbf{f}_{k,m}^{(t)}=\mathbf{A}_{k}^{\mathrm{s},(t)}e^{-j2\pi f_{m}\mathbf{t}_{k}^{(t)}},m=1,2,\cdots,M$ are aligned with the physical directions $\bar{\theta}_{k,m,i}^{(t)},m=1,2,\cdots,M$, which is corresponding to the target user physical directions in $\mathbf{\Psi}_{k}^{i+1}$. This enables that the $t$-th fraction of angular tracking range will be covered. Based on $\mathbf{A}_{k}^{\mathrm{s},(t)}$ and $\mathbf{t}_{k}^{(t)}$, the analog beamformer in the $t$-th time slot $\mathbf{A}_{m}^{(t)}$ is calculated in steps $9$ and $10$.
	
After the analog beamformer $\mathbf{A}_{m}^{(t)}$ has been calculated, the BS transmits training pilot sequence $\mathbf{q}_{k,m}^{(t)}\in\mathcal{C}^{Q\times 1}$ in total $Q$ instants by using $\mathbf{A}_{m}^{(t)}$ in the $t$-th time slot for the $k$-th user at the $m$-th subcarrier. As shown in step $11$, the received signals of the training pilot sequences for $K$ users $\mathbf{Y}_{m,t}\in\mathcal{C}^{K\times Q}$ at subcarrier $m$ can be denoted as

\begin{equation}\label{18}
\mathbf{Y}_{m,t}=\kappa_{m}\mathbf{H}_{m}\mathbf{A}^{(t)}_{m}\mathbf{Q}^{(t)}_{m}+\mathbf{N}^{(t)},
\end{equation}
where $\mathbf{Q}^{(t)}_{m}=[\mathbf{q}_{1,m}^{(t)},\mathbf{q}_{2,m}^{(t)},\cdots,\mathbf{q}_{K,m}^{(t)}]^{H}$ denotes the pilot sequences of all $K$ users with power restriction $\frac{1}{Q}\mathbf{q}_{k,m}^{(t),H}\mathbf{q}_{k,m}^{(t)}=1,k=1,2,\cdots,K$, $\kappa_{m}$ is a normalization coefficient to compensate for the frequency-dependent path gain with $\kappa_{m}=f_\mathrm{c}/f_{m}$ according to (\ref{model2}) which can be realized by digital precoder. To mitigate the inter-user interference caused by the overlapped user tracking range, we suppose the pilot sequences for different users are orthogonal as $\mathbf{q}_{i,m}^{(t),H}\mathbf{q}_{j,m}^{(t)}=0,i\neq j$. Based on the received signals, the user can calculate the label $t_{k}$ and $m_{k}$ of the user physical direction in the next frame in step $13$ by maximizing the power of the product between the received signal and pilot sequence as
\begin{equation}\label{19}
\left(t_{k},m_{k}\right)=\mathop{\arg\max}\limits_{t\in{1,2,\cdots,T},m\in{1,2,\cdots,M}}\|\mathbf{Y}_{m,t,[k,:]}\mathbf{q}_{k,m}^{(t)}\|_{2}^{2}.
\end{equation}
Finally, the BS can obtain the user physical directions $\theta_{k,i+1}^{(0)}$ for all $K$ users through the feedback of $t_{k}$ and $m_{k}$ in step $14$.
	
Three points about the proposed beam zooming based beam tracking scheme should be emphasized. Firstly, since beamforming vectors for different users are generated independently by using different TDs and PSs in the DPP structure, the beam tracking procedures of different users are not separated, but simultaneously carried out. For instance, in the $t$-th time slot, $\mathbf{A}_{m}^{(t)}$ generates required beams for all $K$ users as shown in step $9$ of \textbf{Algorithm 1}. Secondly, the pilot sequence $\mathbf{q}_{k,m}^{(t)}$ for the $k$-th user is generated by the digital precoder $\mathbf{D}_{m}$ and the transmitted signal $\mathbf{s}$. Specifically, the $q$-th pilot of the pilot sequence $\mathbf{q}_{k,m}^{(t)}$ satisfies $[\mathbf{q}_{1,m,[q]}^{(t)},\mathbf{q}_{2,m,[q]}^{(t)},\cdots,\mathbf{q}_{K,m,[q]}^{(t)}]^{T}=\mathbf{D}_{m}\mathbf{s}$, where the digital precoder $\mathbf{D}_{m}$ and the transmitted signal $\mathbf{s}$ can be designed according to the required pilot sequence $\mathbf{q}_{k,m}^{(t)}$, e.g., the pilot sequences $\mathbf{q}_{k,m}^{(t)}$ for different users is designed to be orthogonal with each other in order to mitigate the inter-user interference \cite{Ref:TrackTera2017}. Thirdly, when the user is lost due to blockage, the proposed scheme can also be utilized to quickly discover the user again. Specifically, by setting the parameter in \textbf{Algorithm 1} as $\theta_{k,i}^{(0)}=0$ and $\alpha_{k}=1$, the proposed scheme can realize a whole angle-domain beam tracking, i.e., beam training. In this way, since multiple physical directions are tracked simultaneously in our proposed scheme, the user can be quickly discovered again after blockage by an accelerated beam training procedure.
	
The advantages of the proposed beam zooming based beam tracking scheme mainly lies in two aspects. Firstly, by actively controlling the angular coverage of beams generated by the DPP structure, i.e., the degree of beam split effect, the proposed beam zooming based beam tracking scheme can generate beams aligned with different physical directions simultaneously. Therefore, it can significantly reduce the beam training overhead  compared with the typical beam tracking scheme which only tracks one physical direction in each time slot. Secondly, with the same beam training overhead $T$, the proposed scheme is able to track $M$ times the number of user physical directions as much as the typical beam tracking scheme. This increased number of tracked user physical directions can improve the beam tracking accuracy. Theoretical analysis in Section \ref{Per} and simulation results in Section \ref{Sim} will verify these advantages.

\begin{figure*}
	\centering
		\includegraphics[width=0.83\textwidth]{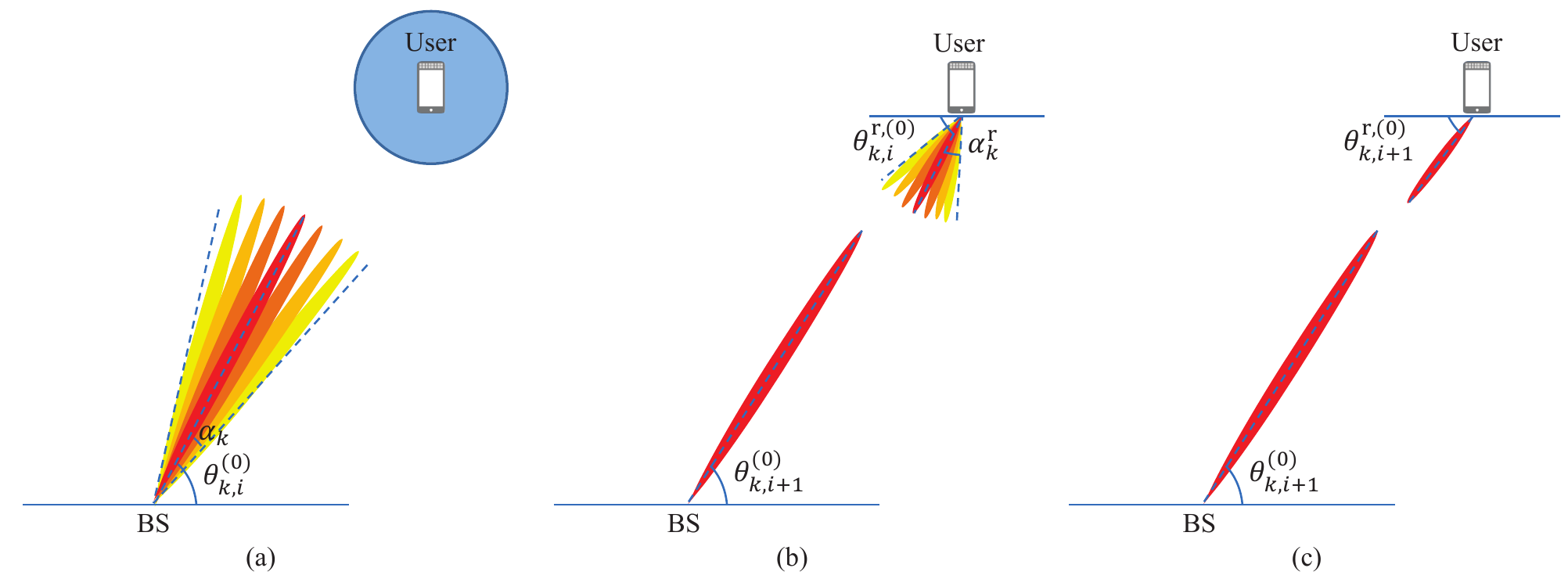}
		\vspace{-2mm}
		\caption{The beam tracking procedure for the proposed beam zooming based scheme in multi-antenna user case: (a) The BS performs beam zooming based beam tracking while user generates omnidirectional beam; (b) The user operates beam zooming based beam tracking while the BS generate beams towards the obtained new physical direction; (c) The BS and the user realize data transmission based on the physical direction obtained from beam tracking.}\vspace{-2mm}
\end{figure*}

\subsection{Extension to Multi-Antenna Users}

In this subsection, we will introduce how to extend our proposed beam zooming based beam tracking scheme into multi-antenna user case. In this case, we consider the $K$ users are each equipped with $N_\mathrm{r}$ antennas. We assume each user uses a single RF chain and employs the DPP structure to mitigate the beam split effect with $K_\mathrm{d}^\mathrm{r}$ TDs.

Under this circumstance, the received signal $\mathbf{y}_{m}$ becomes
\begin{equation}\label{Ex1}
\mathbf{y}_{m}=\mathbf{A}_{m}^\mathrm{r}\mathbf{H}_{m}\mathbf{A}_{m}\mathbf{D}_{m}\mathbf{s}+\mathbf{n},
\end{equation}
where $\mathbf{A}_{m}^\mathrm{r}\in\mathcal{C}^{K\times N_\mathrm{r}K}$ denotes the frequency-dependent analog combining vectors at the user side at subcarrier $m$ with $\mathbf{A}_{m}^\mathrm{r}=\mathrm{blkdiag}\{[\mathbf{f}_{1,m}^\mathrm{r},\mathbf{f}_{2,m}^\mathrm{r},\cdots,\mathbf{f}_{K,m}^\mathrm{r}]^{T}\}$. $\mathbf{f}_{k,m}^\mathrm{r}\in\mathcal{C}^{N_\mathrm{r}\times 1}$ represents the analog combining vector at the $k$-th user, which satisfies $\mathbf{f}_{k,m}^\mathrm{r}=\mathbf{A}^\mathrm{s,r}_{k}e^{-j2\pi f_{m}\mathbf{t}_{k}^\mathrm{r}}$ where $\mathbf{A}_{k}^\mathrm{s,r}\in\mathcal{C}^{N_\mathrm{r}\times K_\mathrm{d}^\mathrm{r}}$ denotes phase shifts provided by PSs with similar form as $\mathbf{A}_\mathrm{k}^\mathrm{s}$ and $\mathbf{t}_{k}^\mathrm{r}\in\mathcal{C}^{K_\mathrm{d}^\mathrm{r}\times 1}$ denotes the time delays provided by TDs. Besides, we should consider the channel model for multi-antenna users, where the channel matrix $\mathbf{H}_{m}$ at the $m$-th subcarrier $\mathbf{H}_{m}=[\mathbf{H}_{1,m}^{T},\mathbf{H}_{2,m}^{T},\cdots,\mathbf{H}_{K,m}^{T}]^{T}\in\mathcal{C}^{N_\mathrm{r}K\times N}$ with $\mathbf{H}_{k,m}\in\mathcal{C}^{N_\mathrm{r}\times N},k=1,2,\cdots,K$ representing the channel of the $k$-th user. Specifically, the channel of the $k$-th user $\mathbf{H}_{k,m}$ can be denoted by the following the ray-based channel model as
\begin{equation}\label{Ex2}
\mathbf{H}_{k,m}=\beta^{(0)}_{k,m}\mathbf{a}_{N_\mathrm{r}}\left(\psi_{k,m}^{\mathrm{r},(0)}\right)\mathbf{a}_{N}^{H}\left(\psi_{k,m}^{(0)}\right),
\end{equation}
where we only consider the LoS path and $\psi_{k,m}^{\mathrm{r},(0)}$ denotes the spatial direction at the user side. Similar to the spatial direction at the BS side $\psi_{k,m}^{(0)}$, $\psi_{k,m}^{\mathrm{r},(0)}=\frac{2d}{c}f_{m}\sin{\tilde{\theta}_{k}^{\mathrm{r},(0)}}$ where $\tilde{\theta}_{k}^{\mathrm{r},(0)}$ is the physical direction at the user side for the LoS path. Here, we also utilize $\theta_{k}^{\mathrm{r},(0)}=\sin{\tilde{\theta}_{k}^{\mathrm{r},(0)}}\in[-1,1]$ to represent the physical direction at the user side.

In the multi-antenna user case with the above signal model, the target of beam tracking changes into tracking the physical directions at both the BS and the user side $(\theta_{k,i+1}^{(0)},\theta_{k,i+1}^{\mathrm{r},(0)})$ at the $(i+1)$-th frame based on the physical directions $(\theta_{k,i}^{(0)},\theta_{k,i}^{\mathrm{r},(0)})$ at the $i$-th frame. Following the thought of the classical beam tracking scheme \cite{Ref:BeamAlignment2010} that the beam tracking procedure at the BS side and the user side can be carried out respectively with the other side fixed, the procedure of the proposed scheme is able to be extended to multi-antenna user case as follows, which is also shown in Fig. 8. Firstly, the BS operates the proposed beam zooming based beam tracking scheme with parameters $\theta_{k,i}^{(0)}$ and $\alpha_{k}$, while the user generates a fixed omnidirectional beam. The omnidirectional beam can be realized by setting zero time delays and using existing design scheme on phase shifts provided by PSs \cite{Ref:Mill2013}. Then, after the BS obtains the new physical direction $\theta_{k,i+1}^{(0)}$, the user performs the proposed scheme with parameters $\theta_{k,i}^{\mathrm{r},(0)}$ and $\alpha_{k}^\mathrm{r}$ (physical direction variation range at the user side), while the BS generate beams towards the physical direction $\theta_{k,i+1}^{(0)}$. Finally, the BS and the user perform beamforming towards the obtained new physical direction $(\theta_{k,i+1}^{(0)},\theta_{k,i+1}^{\mathrm{r},(0)})$ according to (\ref{6}), (\ref{7}) and (\ref{8}) during data transmission.

Since the proposed beam zooming based scheme can realize fast beam tracking by tracking multiple physical directions simultaneously, the beam training overhead of the proposed scheme in multi-antenna user systems would be also relatively small. It should be pointed out that the omnidirectional beam at the user may cause array gain loss. Fortunately, the proposed scheme, which only needs feedback of subcarrier label from the user to the BS, does not require a high data rate. Meanwhile, the BS with much more antennas than the user can provide high array gain in the first stage. Therefore, the omnidirectional beam may have little effect on the proposed scheme in most scenarios. While, to totally overcome this problem, it will be better if a wide beam covering the angle-domain tracking range $[\theta_{k,i}^{\mathrm{r},(0)}-\alpha_{k}^{\mathrm{r}},\theta_{k,i}^{\mathrm{r},(0)}+\alpha_{k}^{\mathrm{r}}]$ can be utilized. These wide beams can be designed by slightly modifying some existing methods\cite{Ref:HierarchicalCode17}. The achievable sum-rate performance of the proposed scheme with multi-antenna users will be verified in Section \ref{Sim}.

\section{Performance Analysis}\label{Per}
In this section, we provide performance analysis of the proposed beam zooming based beam tracking scheme. At first, the lower bound of the required beam training overhead of the proposed scheme is derived. Then, we analyze the achievable sum-rate performance of the proposed scheme. These analysis reveals that the proposed scheme is able to approach the near-optimal achievable sum-rate with very low beam training overhead.
	
\subsection{The Beam Training Overhead}
In the proposed beam zooming based scheme, the frequency-dependent beams generated in each time slot should cover a fraction of the angular tracking range, e.g., $[\theta_{k,i}^{(0)}-\alpha_{k}+\frac{(2t-2)\alpha_{k}}{T},\theta_{k,i}^{(0)}-\alpha_{k}+\frac{2t\alpha_{k}}{T}]$ in the $t$-th time slot. Therefore, in each time slot, the required angular coverage of these frequency-dependent beams is $\frac{\alpha_{k}}{T}$. Note that this angular coverage $\frac{\alpha_{k}}{T}$ cannot be realized for an arbitrary beam training overhead $T$, since the range of physical direction adjustment provided by the DPP structure is limited by the phase shifts provided by TDs, i.e., $\beta_{k,m}\in[-1,1]$. Under this limitation, which has been described in Subsection \ref{FBZ}, when the beam training overhead $T$ is too small, the beam zooming mechanism cannot generate frequency-dependent beams that cover a large angular range $\frac{\alpha_{k}}{T}$. Hence, by considering the limitation of the DPP structure, we prove the lower bound of the required beam training overhead $T_\mathrm{min}$ in this subsection.
	
Specifically, recalling the frequency-dependent phase shifts provided by TDs $\beta_{k,m}$ in \textbf{Lemma 1}, it should satisfy $\beta_{k,m}\in[-1,1]$. Therefore, by substituting the relationship between $\beta_{k,m}$ and $s_{k}$ as $\beta_{k,m}=2(\xi_{m}-1)s_{k}$ in (\ref{12}) in the Appendix A into $\beta_{k,m}\in[-1,1]$, we can obtain that the number of period $s_{k}^{(t)}$ that delayed by TDs for the $k$-th user in the $t$-th time slot should meet
\begin{equation}\label{21}
\left|2(1-\xi_{m})s_{k}^{(t)}\right|\leq 1.
\end{equation}
Note that in the proposed beam zooming based beam tracking scheme, the number of period that delayed by TDs $s_{k}^{(t)}$ is designed as shown in (\ref{16}) and (\ref{17}). Thus, by substituting (\ref{16}) and (\ref{17}) into (\ref{21}), we can obtain the inequality that the beam training overhead $T$ should satisfy
\begin{equation}\label{22}
\left|-P(1-\xi_{m})\left(\bar{\theta}_{k,i,\mathrm{cen}}^{(t)}+(1-\xi_{1})\frac{\alpha_{k}}{T}+\frac{2\xi_{M}\xi_{1}\alpha_{k}}{(\xi_{M}-\xi_{1})T})\right)\right|\leq 1.
\end{equation}
According to step $1$ in \textbf{Algorithm 1}, the central physical direction of the $t$-th fraction of tracking range $\bar{\theta}_{k,i,\mathrm{cen}}^{(t)}$ is $\bar{\theta}_{k,i,\mathrm{cen}}^{(t)}=\theta_{k,i}^{(0)}-\alpha_{k}+\frac{(2t-1)\alpha_{k}}{T}$. Thus, by substituting $\bar{\theta}_{k,i,\mathrm{cen}}^{(t)}=\theta_{k,i}^{(0)}-\alpha_{k}+\frac{(2t-1)\alpha_{k}}{T}$ into (\ref{22}), we obtain
\begin{equation}\label{23}
\left|P(1-\xi_{m})(\theta_{k,i}^{(0)}-\alpha_{k})T+\gamma_{t,m}\alpha_{k}\right|\leq T,
\end{equation}
where we define $\gamma_{t,m}$ as
\begin{equation}\label{24}
\gamma_{t,m}=P(1-\xi_{m})\left(2t-\xi_{1}+\frac{2\xi_{M}\xi_{1}}{\xi_{M}-\xi_{1}}\right),
\end{equation}
in which $t=1,2,\cdots,T$ and $m=1,2,\cdots,M$.
	
From the inequality in (\ref{23}), we can derive the lower bound of the required beam training overhead $T_\mathrm{min}$ for the proposed beam zooming based beam tracking scheme. Specifically, according to (\ref{23}), we know that the training overhead $T$ should satisfy the following two inequalities as
\begin{equation}\label{25}
\begin{aligned}
\left(1-P(1-\xi_{m})(\theta_{k,i}^{(0)}-\alpha_{k})\right)T&\geq\gamma_{t,m}\alpha_{k},\\
\left(1+P(1-\xi_{m})(\theta_{k,i}^{(0)}-\alpha_{k})\right)T&\geq-\gamma_{t,m}\alpha_{k}.
\end{aligned}
\end{equation}
We first consider the case that $m>M/2$. When $m>M/2$, based on (\ref{24}), we can easily know that $\gamma_{t,m}<0,t=1,2,\cdots,T$. In addition, as proved by \textbf{Lemma 3} in Appendix B, $\left(1\pm P(1-\xi_{m})(\theta_{k,i}^{(0)}-\alpha_{k})\right)>0$ holds. Therefore, since $\gamma_{t,m}<0$ and $\left(1\pm P(1-\xi_{m})(\theta_{k,i}^{(0)}-\alpha_{k})\right)>0$, when $m>M/2$, we have
\begin{equation}\label{26}
T>-\frac{\gamma_{t,m}\alpha_{k}}{\left(1+P(1-\xi_{m})(\theta_{k,i}^{(0)}-\alpha_{k})\right)}=\tau_{1}.
\end{equation}
Similarly, for the case $m\leq M/2$, we can obtain
\begin{equation}\label{27}
T>\frac{\gamma_{t,m}\alpha_{k}}{\left(1-P(1-\xi_{m})(\theta_{k,i}^{(0)}-\alpha_{k})\right)}=\tau_{2}.
\end{equation}
Combining (\ref{26}) and (\ref{27}), we know that the lower bound of the require beam training overhead $T_\mathrm{min}$ should be the smallest integer that satisfies (\ref{26}) and (\ref{27}) for any time slot $t$, subcarrier $m$ and physical direction $\theta_{k,i}^{(0)}\in[-1,1]$. Therefore, $T_\mathrm{min}$ can be denoted as
\begin{equation}\label{28}
T_\mathrm{min} =\left\lceil\max\left( \max\limits_{m>M/2,\theta_{k,i}^{(0)},t}\tau_{1},
\max\limits_{m<=M/2,\theta_{k,i}^{(0)},t}\tau_{2}\right)\right\rceil.
\end{equation}

The above analysis provides the lower bound of the required beam training overhead $T_\mathrm{min}$ of the proposed scheme, which is shown in (\ref{28}). We can observe from (\ref{26}), (\ref{27}) and (\ref{28}) that when the angular tracking range (i.e., $\alpha_{k}$) becomes larger or the bandwidth becomes larger (i.e., $\xi_{m}$ becomes larger when $m>M/2$ or $\xi_{m}$ becomes smaller when $m<M/2$), the lower bound of the required beam training overhead $T_\mathrm{min}$ becomes larger. This implies that when the user is moving fast or the system bandwidth is quite large, the proposed scheme requires longer time to track the user. Fortunately, since multiple user physical directions are tracked simultaneously in our proposed scheme, the lower bound of the required beam training overhead $T_\mathrm{min}$ is usually much smaller than that of the typical beam tracking scheme\cite{Ref:BeamAlignment2010}. For example, when $N=256$, $K_\mathrm{d}=16$, $M=128$, $\alpha_{k}=0.1$, $f_{c}=100$ GHz, and $B=10$ GHz, we have $T_\mathrm{min}=2$ from (\ref{28}), which is much smaller than the beam training overhead  $T=52$ of the typical beam tracking scheme \cite{Ref:BeamAlignment2010} as discussed in Subsection \ref{Tra}. Simulation results will also verify this analysis in Section \ref{Sim}.

\subsection{The Achievable Sum-Rate Performance}\label{Rate}
The beam tracking accuracy has a big impact on the achievable sum-rate performance, since the beams cannot be aligned with users when the beam tracking result is inaccurate. In this subsection, we will theoretically establish the relationship between the achievable sum-rate and the beam tracking accuracy. The analysis will illustrate that the proposed scheme is able to achieve the near-optimal sum-rate performance by increasing the number of tracked user physical directions. 
	
Specifically, in the data transmission period, the achievable sum-rate $R$ can be represented as
\begin{equation}\label{26.1}
R=\sum_{1}^{K}\sum_{1}^{M}R_{k,m}=\sum_{1}^{K}\sum_{1}^{M}\log_{2}(1+\zeta_{k,m}),
\end{equation}
where $R_{k,m}$ denotes the achievable rate of the $k$-th user at the $m$-th subcarrier, and $\zeta_{k,m}$ presents the corresponding  signal-to-interference-plus-noise ratio (SINR) as
\begin{equation}\label{27.1}
\zeta_{k,m}=\frac{\left|\mathbf{h}_{k}^{H}\mathbf{A}_{m}\mathbf{D}_{m,[:,k]}\right|^{2}}{\sum_{k^{'}\neq k}^{K}\left|\mathbf{h}_{k}^{H}\mathbf{A}_{m}\mathbf{D}_{m,[:,k']}\right|^{2}+\sigma^{2}}.
\end{equation}
	
We consider high signal-to-noise ratio situation and utilize the classical ZF method\cite{Ref:FundWC2005}  to design the digital precoder $\mathbf{D}_{m}$ as
\begin{equation}
\mathbf{D}_{m}=\mathbf{H}_{m,\mathrm{eq}}^{H}\left(\mathbf{H}_{m,\mathrm{eq}}\mathbf{H}_{m,\mathrm{eq}}^{H}\right)^{-1}\mathbf{\Gamma},
\end{equation}
where the power normalization matrix $\mathbf{\Gamma}$ is a diagonal matrix, whose diagonal elements are designed to satisfy the transmission power constraint $\|\mathbf{A}_{m}\mathbf{D}_{m,[:,k]}\|_{2}^{2}=\rho$. Thus, the power normalization matrix $\mathbf{\Gamma}$ satisfies
\begin{equation}\label{29}
\begin{aligned}
\mathbf{\Gamma}_{[k,:]}&\left(\mathbf{H}_{m,\mathrm{eq}}^{H}\mathbf{H}_{m,\mathrm{eq}}\right)^{-1}\mathbf{H}_{m,\mathrm{eq}}\mathbf{A}_{m}^{H}\\&\times\mathbf{A}_{m}\mathbf{H}_{m,\mathrm{eq}}^{H}\left(\mathbf{H}_{m,\mathrm{eq}}\mathbf{H}_{m,\mathrm{eq}}^{H}\right)^{-1}\mathbf{\Gamma}_{[:,k]}=\rho.
\end{aligned}
\end{equation}
Note that in the data transmission period, the analog beamformer $\mathbf{A}_{m}$ generates beams towards user physical directions $\theta_{k}^{(0)},k=1,2,\cdots,K$, which are designed according to (\ref{6}), (\ref{7}) and (\ref{8}). Without loss of generality, we assume that the user physical directions $\theta_{k}^{(0)},k=1,2,\cdots,K$ for different users are resolvable in the angle domain, which means that the users are served by different beams in the same frame. This assumption is reasonable due to the narrow beam width in THz massive MIMO systems. Thus, by utilizing the approximate orthogonality of different beamforming vectors \cite{Ref:Aodcodebook2018}, we have
\begin{equation}\label{30}
\mathbf{A}_{m}^{H}\mathbf{A}_{m}\approx\mathbf{I}_{K}.
\end{equation}
Substituting (\ref{30}) and $\mathbf{H}_{m,\mathrm{eq}}=\mathbf{H}_{m}\mathbf{A}_{m}$ into (\ref{29}), we have
\begin{equation}\label{31}
\mathbf{\Gamma}_{[k,:]}\left(\mathbf{A}_{m}^{H}\mathbf{H}_{m}^{H}\mathbf{H}_{m}\mathbf{A}_{m}\right)^{-1}\mathbf{\Gamma}_{[:,k]}=\rho.
\end{equation}
Since THz communications heavily rely on the dominant LoS path, we consider the case that the channel only contains a single LoS path, i.e., $L=1$. Thus, the channel $\mathbf{H}_{m}$ satisfies $\mathbf{H}_{m}=[\mathbf{h}_{1,m}^{T},\mathbf{h}_{2,m}^{T},\cdots,\mathbf{h}_{K,m}^{T}]^{T}$ and $\mathbf{h}_{k,m}=\beta^{(0)}_{k,m}\mathbf{a}_{N}^{H}\left(\psi_{k,m}^{(0)}\right)$ according to (\ref{1}). In addition, the analog beamformer $\mathbf{A}_{m}$ consists of beamforming vectors as $\mathbf{A}_{m}=[\mathbf{f}_{1,m},\mathbf{f}_{2,m},\cdots,\mathbf{f}_{K,m}]$. Considering the array gain achieved by the beamforming vector $\mathbf{f}_{k,m}$ on the user physical direction $\theta_{k}^{(0)}$ at the $m$-th subcarrier frequency $f_{m}$ is defined as $\eta(\mathbf{f}_{k,m},\theta_{k}^{(0)},f_{m})=|\mathbf{a}_{N}^{H}(\psi_{k,m}^{(0)})\mathbf{f}_{k,m}|$, the elements of the matrix $\mathbf{H}_{m}\mathbf{A}_{m}$ satisfy
\begin{equation}\label{32}
\begin{aligned}
|(\mathbf{H}_{m}\mathbf{A}_{m})_{[u,v]}|&=|\beta_{u,m}^{(0)}\mathbf{a}_{N}^{H}(\psi_{u,m}^{(0)})\mathbf{f}_{v,m}|\\
&=|\beta_{u,m}^{(0)}\eta(\mathbf{f}_{v,m},\theta_{u}^{(0)},f_{m})|,
\end{aligned}
\end{equation}
where $\mathbf{f}_{v,m}=\mathbf{A}_{m,[:,v]}$ denotes the frequency-dependent beamforming vector, which generates a beam aligned with the user physical direction $\theta_{v}^{(0)}$ for the $v$-th user, and $u,v=1,2,\cdots,K$. Since we assume that users are separable in the angle domain, the beamforming vector for one user achieves the near-zero array gain on the other users' physical directions due to the narrow beamwidth. Thus, we have the array gain $\eta(\mathbf{f}_{v,m},\theta_{u}^{(0)},f_{m})\approx0$ when $v\neq u$, which leads to $(\mathbf{H}_{m}\mathbf{A}_{m})_{[u,v]}\approx0$ when $v\neq u$ according to (\ref{32}). Hence, we know that the matrix $\mathbf{H}_{m}\mathbf{A}_{m}$ is a diagonal matrix with diagonal elements satisfying $|(\mathbf{H}_{m}^{H}\mathbf{A}_{m})_{[k,k]}|=|(\beta_{k,m}^{(0)})\eta(\mathbf{f}_{k,m},\theta_{u}^{(0)},f_{m})|$ for $k=1,2,\cdots,K$. Then, by substituting the diagonal elements of the matrix $\mathbf{H}_{m}\mathbf{A}_{m}$ into (\ref{31}), we can obtain the diagonal elements of the power normalization matrix $\mathbf{\Gamma}$ as
\begin{equation}\label{33}
\mathbf{\Gamma}_{[k,k]}=\sqrt{\eta(\mathbf{f}_{k,m},\theta_{k}^{(0)},f_{m})^{2}(\beta_{k,m}^{(0)})^{2}\rho}.
\end{equation}
Based on (\ref{33}) and the zero-forcing property of the digital precoder $\mathbf{D}_{m}$, the achievable rate $R_{k,m}$ of the $k$-th user at the $m$-th subcarrier defined in (\ref{26.1}) can be rewritten as
\begin{equation}\label{34}
R_{k,m}=\log_{2}\left(1+\frac{\rho}{\sigma^{2}}(\beta_{k,m}^{(0)})^{2}\eta^{2}(\mathbf{f}_{k,m},\theta_{k}^{(0)},f_{m})\right).
\end{equation}
Based on (\ref{34}), we know that the achievable rate $R_{k,m}$ in the $i$-th frame with user physical directions $\theta_{k,i}^{(0)},k=1,2,\cdots,K$ can be denoted as
\begin{equation}\label{34-1}
R_{k,m}=\log_{2}\left(1+\frac{\rho}{\sigma^{2}}(\beta_{k,m}^{(0)})^{2}\eta^{2}(\mathbf{f}_{k,m},\theta_{k,i}^{(0)},f_{m})\right).
\end{equation}
We can observe from (\ref{34-1}) that the achievable rate $R_{k,m}$ is mainly determined by the array gain achieved by the beamforming vector $\mathbf{f}_{k,m}$ on the $k$-th user's physical direction $\theta_{k,i}^{(0)}$ in the $i$-th frame. Let $\hat{\theta}_{k,i}^{(0)}$ denote the beam tracking results in the $i$-th frame. Note that after the physical directions $\hat{\theta}_{k,i}^{(0)}$ has been obtained through beam tracking, the beamforming vectors  $\mathbf{f}_{k,m}$ at all $M$ subcarriers are designed to generate beams aligned with $\hat{\theta}_{k,i}^{(0)}$ in the data transmission period. Therefore, as proved in (\ref{App2}) in \textbf{Lemma 1}, by substituting $\beta_{k,m}=2(\xi_{m}-1)s_{k}$ in (\ref{12}) in the Appendix A and $s_{k}=(P\hat{\theta}_{k,i}^{(0)})/2$ in (\ref{8})  into (\ref{App2}), the array gain $\eta(\mathbf{f}_{k,m},\theta_{k,i}^{(0)},f_{m})$ can be further expressed as
\begin{equation}\label{35}
\begin{aligned}
\eta(\mathbf{f}_{k,m},\theta_{k,i}^{(0)},f_{m})=|\frac{1}{N}\Xi_{K_\mathrm{d}}&(\xi_{m}P(\hat{\theta}_{k,i}^{(0)}-\theta_{k,i}^{(0)}))\\	&\times\Xi_{P}(\hat{\theta}_{k,i}^{(0)}-\xi_{m}\theta_{k,i}^{(0)})|.
\end{aligned}
\end{equation}
Assuming the correct subcarrier can be detected and considering that $\hat{\theta}_{k,i}^{(0)}$ is chosen from the target physical direction set $\mathbf{\Psi}_{k}^{i}$ as shown in step $3$ of \textbf{Algorithm 1}, the beam tracking result $\hat{\theta}_{k,i}^{(0)}$ can be seen as a quantization of the user physical direction $\theta_{k,i}^{(0)}$. Recalling the elements in the targeted physical direction set $\mathbf{\Psi}_{k}^{i}$  defined in (\ref{15}), we know that the quantization error $\left|\hat{\theta}_{k,i}^{(0)}-\theta_{k,i}^{(0)}\right|$ satisfies
\begin{equation}\label{36}
\begin{aligned}
\left|\hat{\theta}_{k,i}^{(0)}-\theta_{k,i}^{(0)}\right|&\leq\arg\max_{m,t}\frac{1}{2}\left(\bar{\theta}_{k,m+1,i}^{(t)}-\bar{\theta}_{k,m,i}^{(t)}\right)\\
&\overset{(a)}{\leq}\arg\max_{m}\frac{\xi_{M}\xi_{1}}{\xi_{M}-\xi_{1}}\left(\frac{1}{\xi_{m}}-\frac{1}{\xi_{m+1}}\right)\frac{\alpha_{k}}{T},
\end{aligned}
\end{equation}
where $(a)$ holds due to the definition of $\bar{\theta}_{k,m,i}^{(t)}$ in (\ref{15}). We can observe from (\ref{36}) that the maximum value of the right side of (\ref{36}) can be achieved when $m=1$. Thus, $\left|\hat{\theta}_{k,i}^{(0)}-\theta_{k,i}^{(0)}\right|$ achieves its maximum value when $m=1$, which means
\begin{equation}\label{37}
\left|\hat{\theta}_{k,i}^{(0)}-\theta_{k,i}^{(0)}\right|\leq\frac{\xi_{M}\xi_{1}}{\xi_{M}-\xi_{1}}\left(\frac{1}{\xi_{1}}-\frac{1}{\xi_{2}}\right)\frac{\alpha_{k}}{T}\overset{(a)}{=}\zeta\frac{\alpha_{k}}{TM},
\end{equation}
where (a) comes from $f_{m}=f_{c}+\frac{B}{M}(m-1-\frac{M-1}{2})$, and we define $\zeta=\frac{\xi_{M}}{\xi_{1}+B/(f_{c}M)}$. Then, based on (\ref{37}), we can bound the array gain by substituting (\ref{37}) into (\ref{35}) as
\begin{equation}\label{38}
\begin{aligned}
\eta^{2}&(\mathbf{f}_{k,m},\theta_{k,i}^{(0)},f_{m})\geq\frac{1}{N^{2}}\Xi_{K_\mathrm{d}}^{2}\left(\frac{\zeta\xi_{m}P\alpha_{k}}{TM}\right)\Xi_{P}^{2}(\hat{\theta}_{k,i}^{(0)}-\xi_{m}\theta_{k,i}^{(0)})\\
&\overset{(a)}\geq\frac{1}{N^{2}}\Xi_{K_\mathrm{d}}^{2}\left(\frac{\zeta\xi_{m}P\alpha_{k}}{TM}\right)\Xi_{P}^{2}\left((1-\xi_{M})\theta_{k,i}^{(0)}+\zeta\frac{\alpha_{k}}{TM}\right),
\end{aligned}
\end{equation}
where $(a)$ comes from $|\hat{\theta}_{k,i}^{(0)}-\xi_{m}\theta_{k,i}^{(0)}|\geq\left((1-\xi_{M})\theta_{k,i}^{(0)}+\zeta\frac{\alpha_{k}}{TM}\right)$, which can be proved by (\ref{37}) and $\xi_{m}\leq\xi_{M}$. Thus, based on (\ref{26.1}) and (\ref{38}), the achievable rate $R_{k,m}$ of the $k$-th user at the $m$-th subcarrier can be lower bounded as
\begin{equation}\label{39}
\begin{aligned}
R_{k,m}&\geq R_{k,m}^\mathrm{bound}=\log_{2}\left(1+\frac{\rho(\beta_{k,m}^{(0)})^{2}}{\sigma^{2}}\eta_\mathrm{min}\right),
\end{aligned}
\end{equation}
where
\begin{equation}\label{40}
\eta_\mathrm{min}=\frac{1}{N^{2}}\Xi_{K_\mathrm{d}}^{2}\left(\frac{\zeta\xi_{m}P\alpha_{k}}{TM}\right)\Xi_{P}^{2}\left((1-\xi_{M})\theta_{k,i}^{(0)}+\zeta\frac{\alpha_{k}}{TM}\right).
\end{equation}
Finally, we can obtain the lower bound of the achievable sum-rate as $R\geq\sum_{k=1}^{K}\sum_{m=1}^{M}R_{k,m}^\mathrm{bound}$. 
	
The lower bound in (\ref{39}) indicates that the beam tracking accuracy is vital for the achievable sum-rate performance. Specifically, we can observe from (\ref{39}) and (\ref{40}) that the lower bound of $R_{k,m}$ is mainly decided by the maximum quantization error $\zeta\frac{\alpha_{k}}{TM}$. As the maximum quantization error $\zeta\frac{\alpha_{k}}{TM}$ becomes larger, the variables $|\zeta\frac{\xi_{m}P\alpha_{k}}{TM}|$ and $|(1-\xi_{M})\theta_{k,i}^{(0)}+\zeta\frac{\alpha_{k}}{TM}|$  of Dirichlet Functions diverge from $0$, which decreases $R_{k,m}$ due to the power-focusing property of the Direchlet Function\cite{Ref:PhasesArray2005}. Therefore, a lower quantization error, which indicates a higher beam tracking accuracy, will result in a better achievable sum-rate performance. Since $M$ times the number of user physical directions are tracked simutaneously, the proposed beam zooming based beam tracking scheme can realize much smaller quantization error (i.e., $\zeta\frac{\alpha_{k}}{TM}$) than that realized by the typical beam tracking scheme \cite{Ref:BeamAlignment2010} (i.e., $\frac{\alpha_{k}}{T}$). Consequently, the proposed scheme is able to achieve better achievable sum-rate performance. Moreover, since the number of subcarriers $M$ is usually large, e.g., $M=128$, the maximum quantization error $\zeta\frac{\alpha_{k}}{TM}$ in our proposed scheme is usually close to zero. For example, when $M=128$, $\alpha_{k}=0.1$, $f_{c}=100$ GHz, $B=10$ GHz, and $T=2$,  we have $\zeta\frac{\alpha_{k}}{TM}=4.3\times 10^{-4}$. Due to such a small maximum quantization error, the proposed scheme is expected to achieve the near-optimal achievable sum-rate performance. 

It should be pointed out that the above analysis is based on the assumption that the proposed scheme can correctly detect the accurate subcarrier. While, in practical systems, due to the noise, small tracking range and the overlap between beams at different subcarriers, the detection error will occur. Fortunately, as shown in Fig. 6, thanks to the power-focusing property and approximate orthogonality of beams at different subcarriers, the detection error will almost only occur between adjacent subcarriers. Under this circumstance, the error of physical direction $\left|\hat{\theta}_{k,i}^{(0)}-\theta_{k,i}^{(0)}\right|$ is quite small and can still be bounded with a great probability by $\bar{\zeta}\frac{\alpha_{k}}{TM}$ where $\bar{\zeta}$ is small (e.g., $\bar{\zeta}= 4$ in simulation results in Fig. 10). Therefore, according to the analysis in (\ref{37})-(\ref{40}), the near-optimal achievable sum-rate can still be expected even though the detection error on subcarriers is taken into consideration. The above analysis will be verified by simulation results in Section \ref{Sim}.

\section{Simulation Results}\label{Sim}
\begin{figure}
	\centering
	\includegraphics[width=0.46\textwidth]{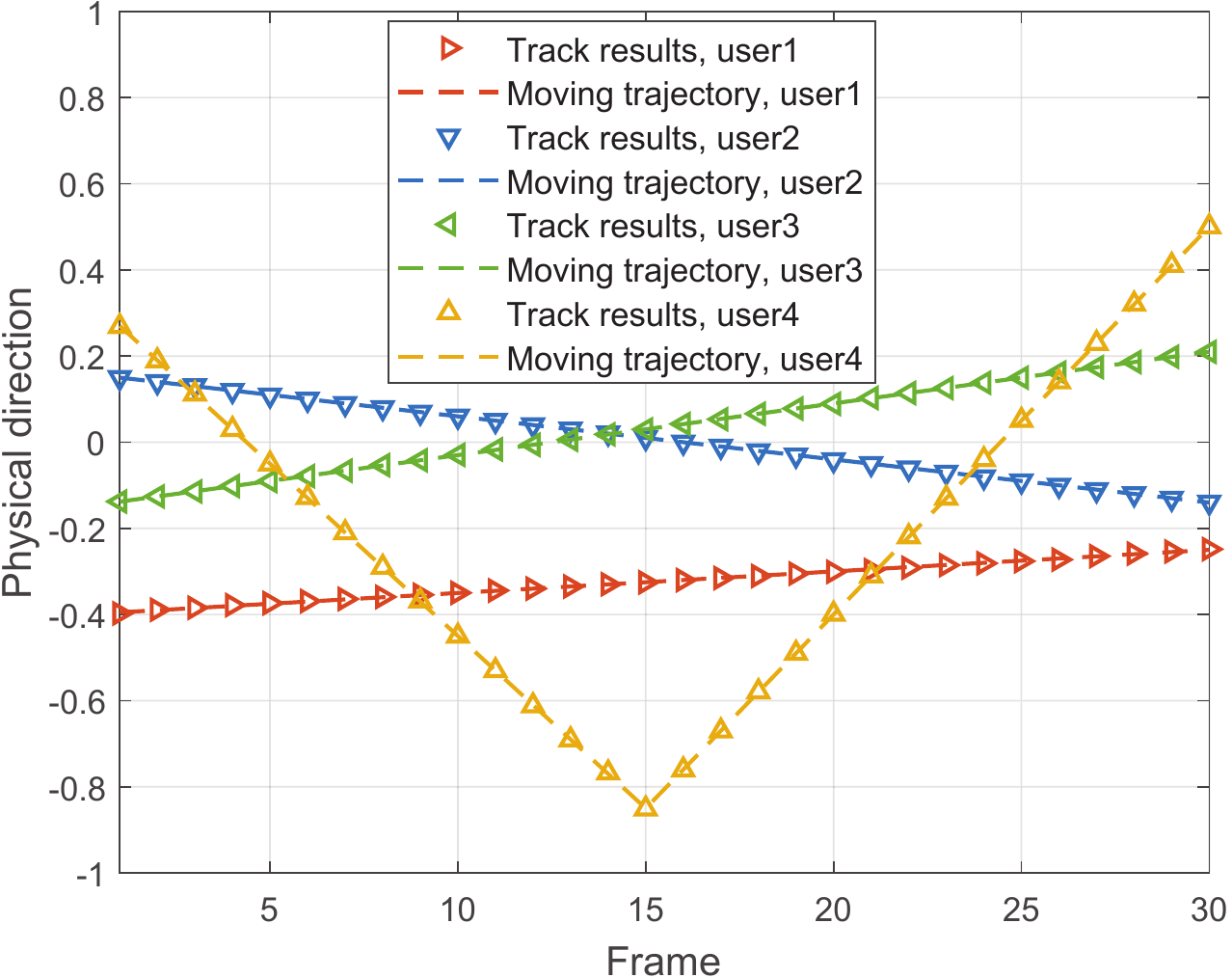}
	\caption{The beam tracking accuracy of the proposed beam zooming based beam tracking scheme.}
\end{figure}
	
In this section, extensive simulation results are provided to show the performance of the proposed beam zooming based beam tracking scheme. The parameters of the THz massive MIMO system with the DPP structure are set as: $N=256$, $M=128$, $K=4$, $K_\mathrm{d}=16$, $f_{c}=100$ GHz and $B=10$ GHz, where the value of $K_\mathrm{d}$ is chosen according to the criterion in \cite{Ref:DPP2019}. For the THz channel, considering that the THz signals are quasi-optical, we set $L=1$. The path gain is considered to be frequency-dependent with the model in (\ref{model2}). We assume the path gain at the central frequency $g_{k,\mathrm{c}}$ as the reference path gain with $g_{k,\mathrm{c}}=1$ without loss of generality. The length of training pilot sequence is $Q=10$. The physical directions of different users $\theta_{k,i}^{(0)}$ are randomly generated by the distribution following $\mathcal{U}(-1,1)$, and the angular variation range $\alpha_{k}$ of user physical directions follows $\mathcal{U}(0,\alpha_{k,\mathrm{max}})$ where $\alpha_{k,\mathrm{max}}$ is the maximum variation range of the user physical direction for the $k$-th user. In addition, due to the low-dimensional property of the equivalent channel $\mathbf{H}_{m,\mathrm{eq}}$, we assume that this equivalent channel can be reliably estimated in the equivalent channel estimation period as shown in Fig. 3, which can be easily realized by using traditional channel estimation methods, e.g., the least square method \cite{Ref:mmWaveMIMO2016}. The analog beamformer $\mathbf{A}_{m}$ is designed based on the physical directions obtained from the proposed beam tracking scheme, according to (\ref{6}), (\ref{7}) and (\ref{8}). Considering THz communications tend to be noise-dominant, the digital precoder $\mathbf{D}_{m}$ is designed by minimum mean square error (MMSE) precoding based on the equivalent channel to mitigate the inter-user interferences\cite{Ref:MMSEPre2019}. The achievable sum-rate performance is calculated by using (\ref{26.1}) and (\ref{27.1}), and the signal-to-noise ratio is defined as $\frac{\rho}{\sigma^{2}}$.
	
Fig. 9 shows the beam tracking accuracy of the proposed  scheme, where the beam tracking results of four users in $30$ consecutive frames are shown, and we set SNR $=10$ dB, $\alpha_{k}=0.1$ and $T=5$. The initial user physical direction $\theta_{k,0}^{(0)}$ and the moving velocity $\theta_{k,i+1}^{(0)}-\theta_{k,i}^{(0)}$ of the users are set as $[\theta_{1,0}^{(0)},\theta_{2,0}^{(0)},\theta_{3,0}^{(0)},\theta_{4,0}^{(0)}]=[-0.4,0.16,-0.15,0.35]$, $\theta_{1,i+1}^{(0)}-\theta_{1,i}^{(0)}=0.005$, $\theta_{2,i+1}^{(0)}-\theta_{1,i}^{(0)}=-0.01$,
$\theta_{3,i+1}^{(0)}-\theta_{1,i}^{(0)}=0.012$,
$\theta_{4,i+1}^{(0)}-\theta_{1,i}^{(0)}=-0.08$ for $i<15$, and $\theta_{4,i+1}^{(0)}-\theta_{1,i}^{(0)}=0.08$ for $i\geq15$, respectively. We can observe from Fig. 9 that the proposed scheme can correctly track the user trajectory. The very small deviation between the actual physical directions and the tracked physical directions is negligible, even for the user 4 who changes its direction in motion.

\begin{figure}
	\centering
		\includegraphics[width=0.44\textwidth]{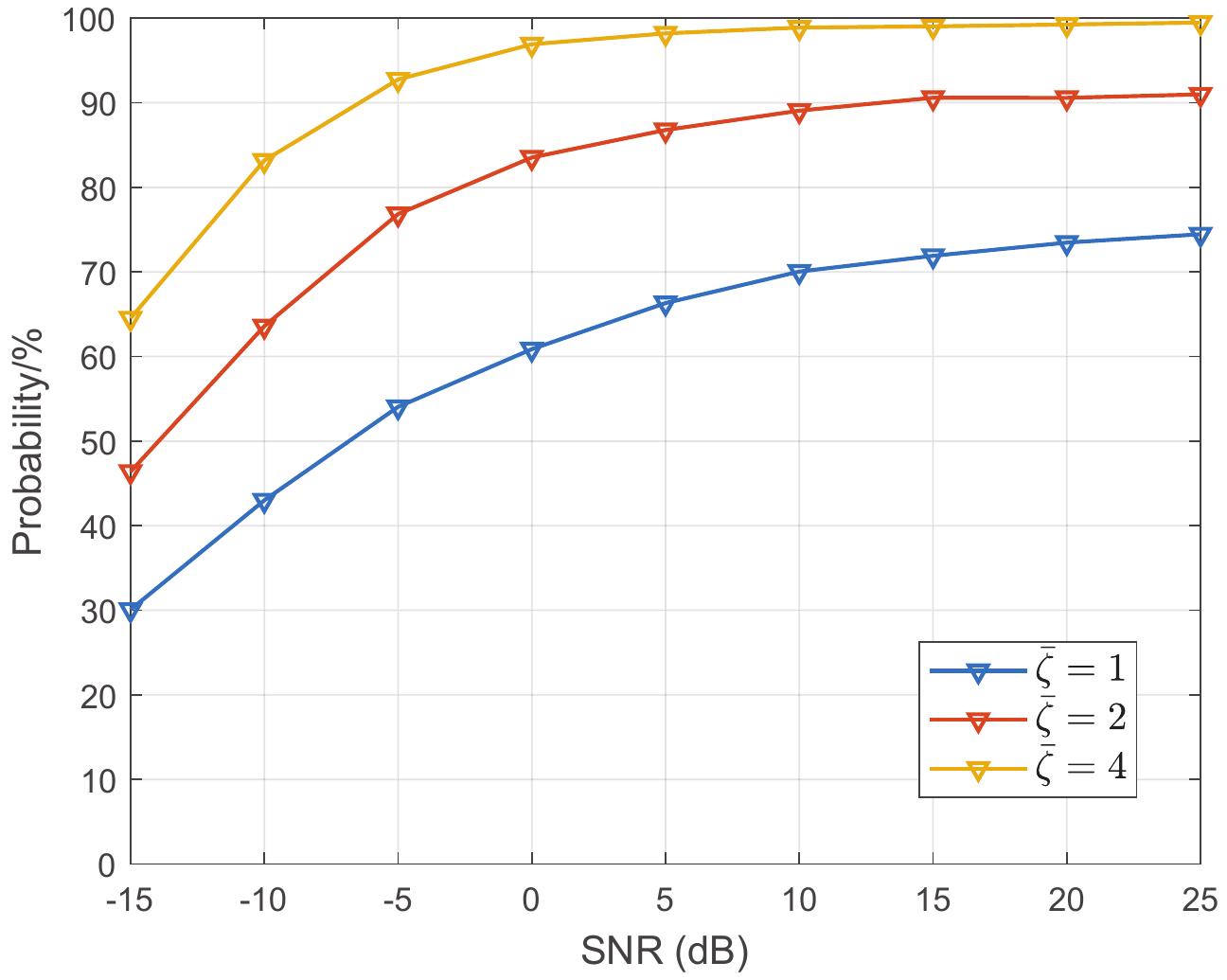}
		\caption{The probability that the estimation error of the physical direction satisfies $\left|\hat{\theta}_{k,i}^{(0)}-\theta_{k,i}^{(0)}\right|\leq \bar{\zeta}\frac{\alpha}{TM}$ with $\bar{\zeta}=1$, $\bar{\zeta}=2$, and $\bar{\zeta}=4$.}
\end{figure}

Fig. 10 illustrates the tracking accuracy with $\alpha_{k,\mathrm{max}}=0.1$ and $T=2$. Specifically, we show the probability that the error of physical direction $\left|\hat{\theta}_{k,i}^{(0)}-\theta_{k,i}^{(0)}\right|$ satisfies $\left|\hat{\theta}_{k,i}^{(0)}-\theta_{k,i}^{(0)}\right|\leq \bar{\zeta}\frac{\alpha_{k}}{TM}$ against SNR level. In Fig. 10, three cases of $\bar{\zeta}$ is considered with $\bar{\zeta}=1$, $\bar{\zeta}=2$ and $\bar{\zeta}=4$. Fig. 10 shows that when SNR is larger than $10$ dB, the probability that $\left|\hat{\theta}_{k,i}^{(0)}-\theta_{k,i}^{(0)}\right|\leq 4\frac{\alpha_{k}}{TM}$ holds is larger than $99\%$. This indicates the error of physical direction is bounded with a great probability by $4\frac{\alpha_{k}}{TM}=0.0016$, which is quite small due to a large number of subcarriers $M$. Consequently, we can conclude that the proposed scheme can realize accurate physical direction estimation.

\begin{figure}
	\centering
	\includegraphics[width=0.44\textwidth]{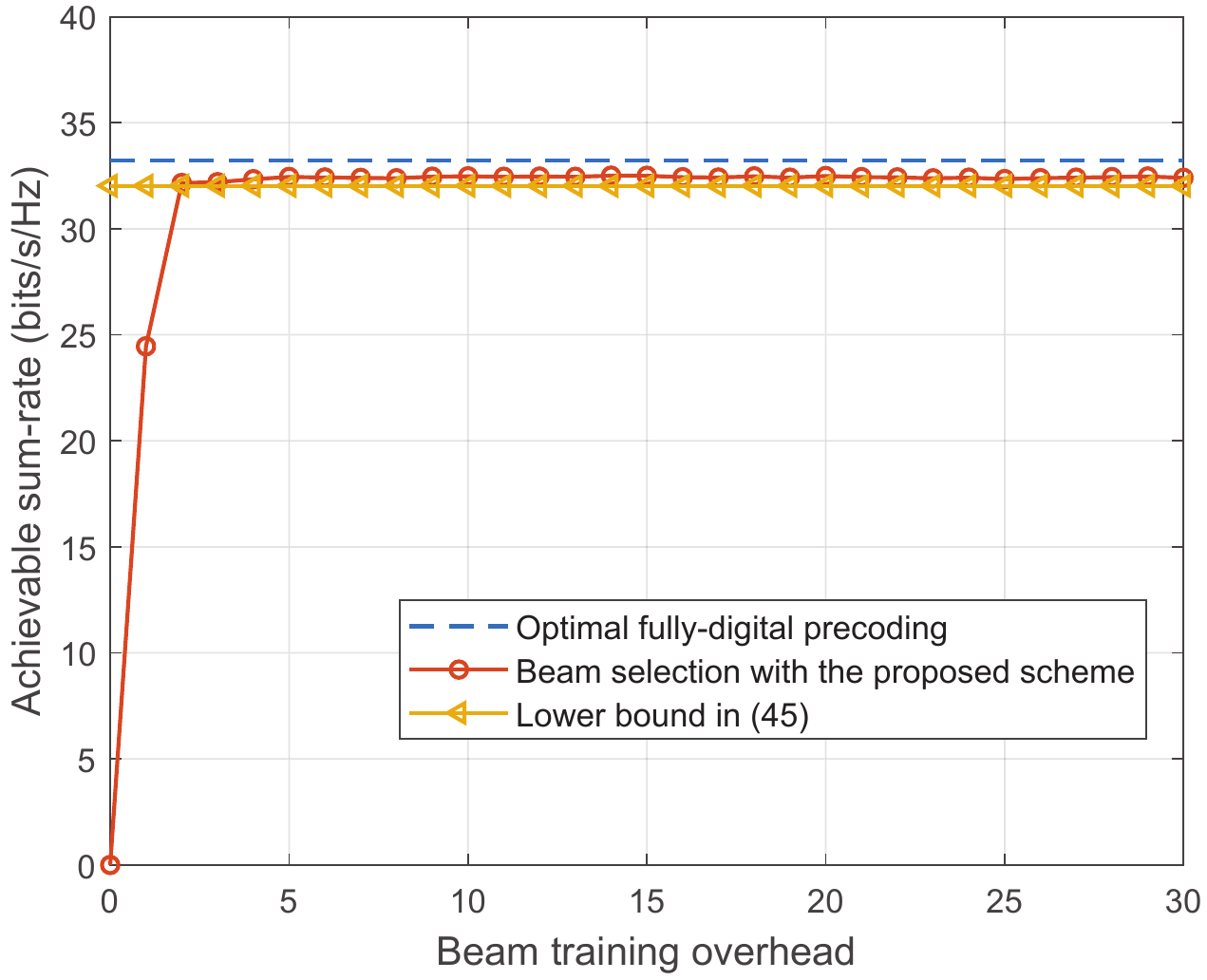}
	\caption{The achievable sum-rate performance against the beam training overhead $T$.}
\end{figure}

Fig. 11 compares the achievable sum-rate performance against the beam training overhead $T$ between the optimal fully-digital MMSE precoding\cite{Ref:MMSEPre2019} with perfect channel information, the beam selection based precoding method with physical directions tracked by the proposed scheme, and the lower bound derived in (\ref{39}). We set $\alpha_{k,\mathrm{max}}=0.1$ for users and SNR = $30$ dB. From Fig. 11, we can observe that by utilizing the proposed beam zooming based beam tracking scheme, the beam selection based precoding method can almost approach the achievable sum-rate performance of the optimal fully-digital precoding. For example, when $T>2$, the beam selection based precoding method with the proposed scheme can achieve nearly $98\%$ achievable sum-rate performance of the optimal fully-digital precoding, which reveals that the proposed scheme can realize accurate beam tracking. In addition, we can find that the achievable sum-rate performance of the beam selection based precoding method with the proposed scheme is tightly lower-bounded by (\ref{39}), which verifies the performance analysis in Subsection \ref{Rate}.
	
\begin{figure}
	\centering
	\includegraphics[width=0.46\textwidth]{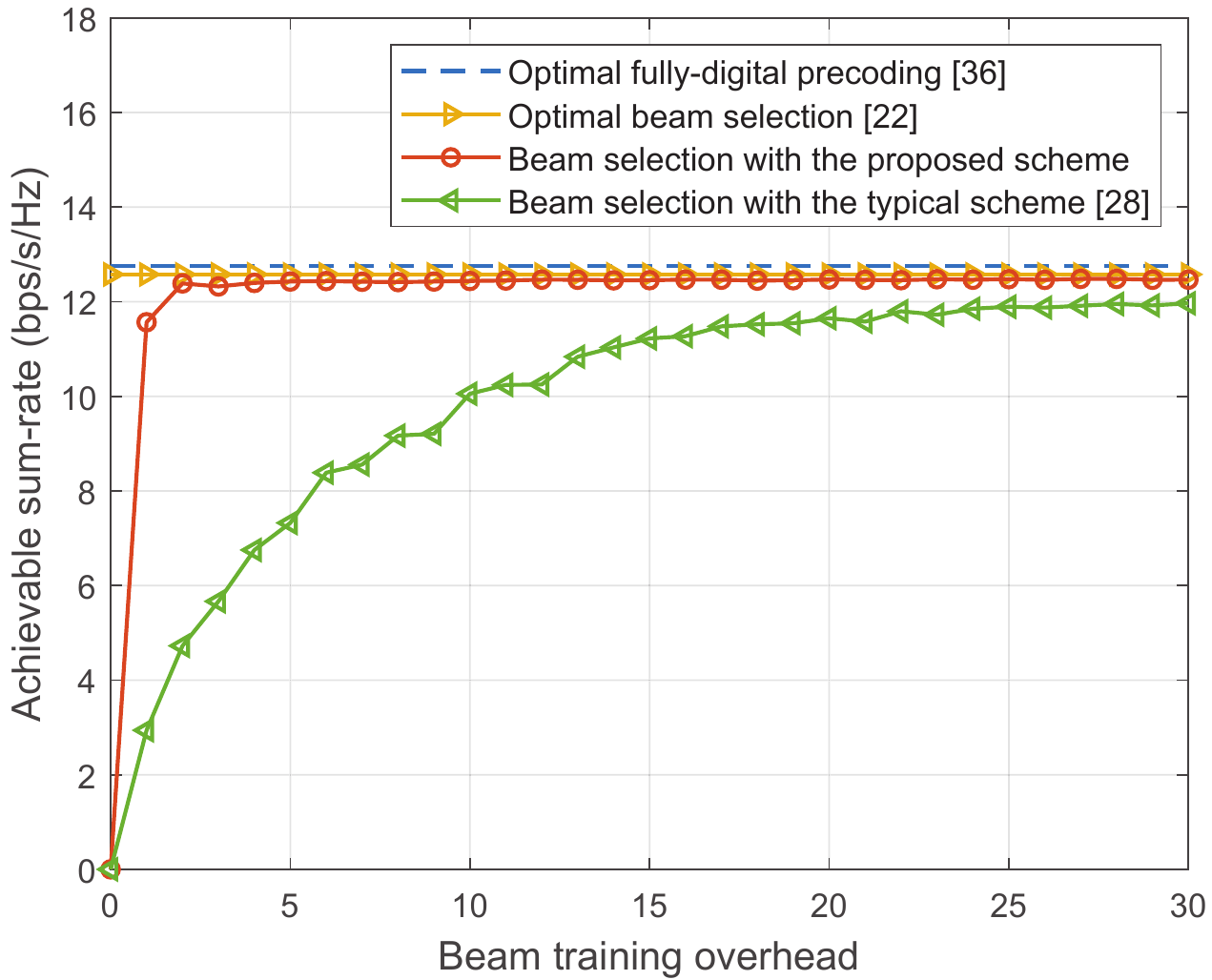}
	\caption{The achievable sum-rate performance against the beam training overhead $T$ with $\alpha_{k,\mathrm{max}}=0.1$.}
\end{figure}
	
\begin{figure}
	\centering
	\includegraphics[width=0.46\textwidth]{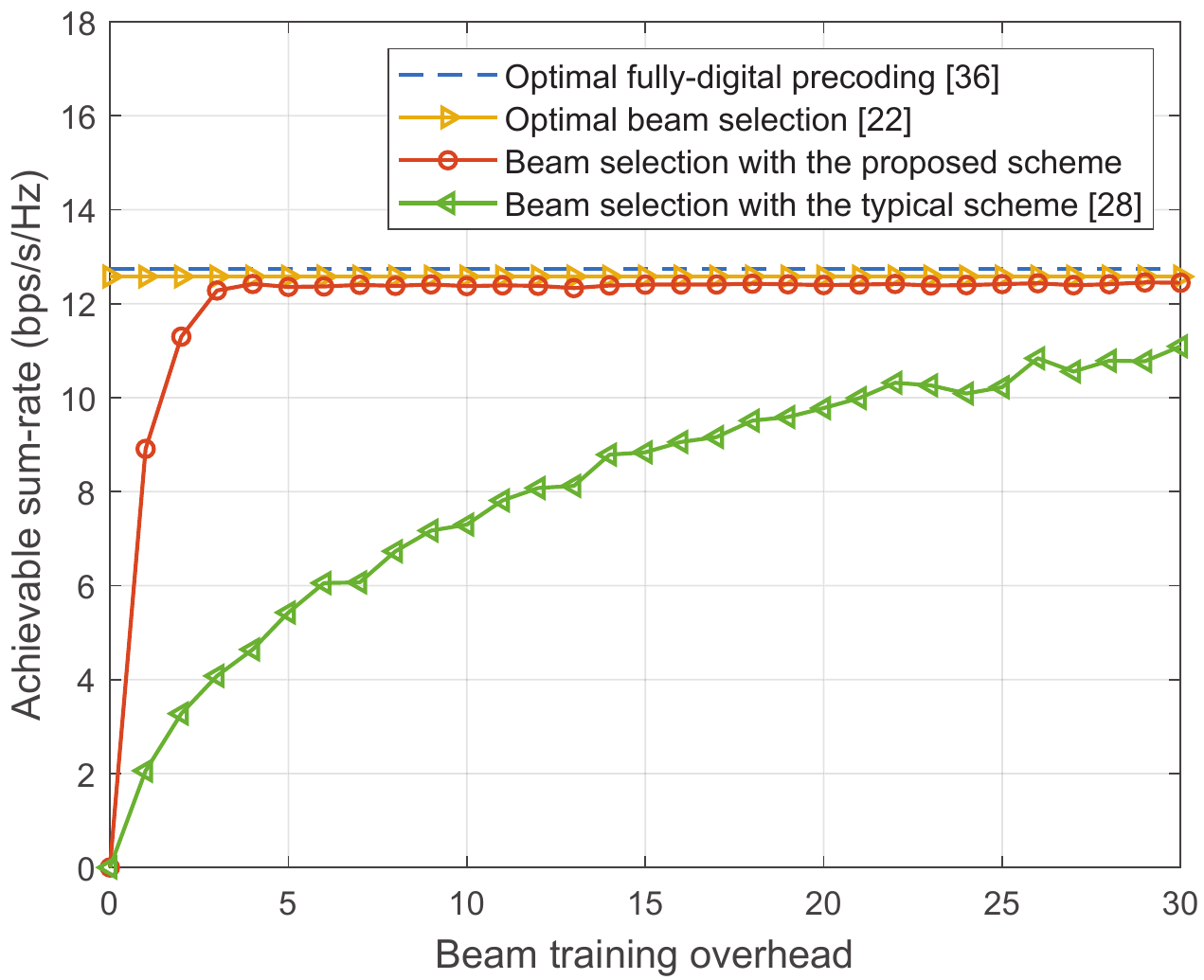}
	\caption{The achievable sum-rate performance against the beam training overhead $T$ with $\alpha_{k,\mathrm{max}}=0.2$.}
\end{figure}
	
In Fig. 12 and Fig. 13, we provide the achievable sum-rate performance against the beam training overhead $T$, where the following four schemes are compared: the optimal fully-digital MMSE precoding\cite{Ref:MMSEPre2019} with perfect channel information, beam selection based precoding method \cite{Ref:BeamSe2018} with the perfect physical directions, beam selection based precoding method with the physical directions tracked by the proposed scheme, and beam selection based precoding method with the physical directions tracked by the typical beam tracking scheme \cite{Ref:BeamAlignment2010} as described in Subsection \ref{Tra}. The cases with different variation range of user physical direction are set as $\alpha_{k,\mathrm{max}}=0.1$ in Fig. 12 and $\alpha_{k,\mathrm{max}}=0.2$ in Fig. 13. We can observe from Fig. 12 and Fig. 13 that by exploiting the proposed beam zooming based beam tracking scheme, the beam selection based precoding method can achieve near-optimal achievable sum-rate performance in both $\alpha_{k,\mathrm{max}}=0.1$ and $\alpha_{k,\mathrm{max}}=0.2$ cases with low training overhead, which is consistent with the theoretical analysis in Section \ref{Per}. For instance, when $T>2$ with $\alpha_{k,\mathrm{max}}=0.1$ or $T>4$ with $\alpha_{k,\mathrm{max}}=0.2$, the achievable sum-rate gap between the beam selection based precoding method with the perfect physical directions and that with the physical directions tracked by the proposed scheme is negligible. Besides, we can observe that the proposed  scheme outperforms the typical beam tracking scheme in two aspects. Firstly, to achieve the same achievable sum-rate performance, the required training overhead of the proposed beam zooming based beam tracking scheme is much smaller than that of the typical scheme. For example, for the achievable sum-rate $10$ bits/s/Hz, the proposed scheme can reduce the beam tracking overhead by about $90$\%. Secondly, with the same beam training overhead, the beam zooming based beam tracking scheme achieves better achievable sum-rate performance than the typical beam tracking scheme. The improvement above is mainly because that through the active control of the degree of the beam split effect, the proposed scheme can track multiple potential user physical directions simultaneously and increase the number of tracked user physical directions.

\begin{figure}
	\centering
	\includegraphics[width=0.46\textwidth]{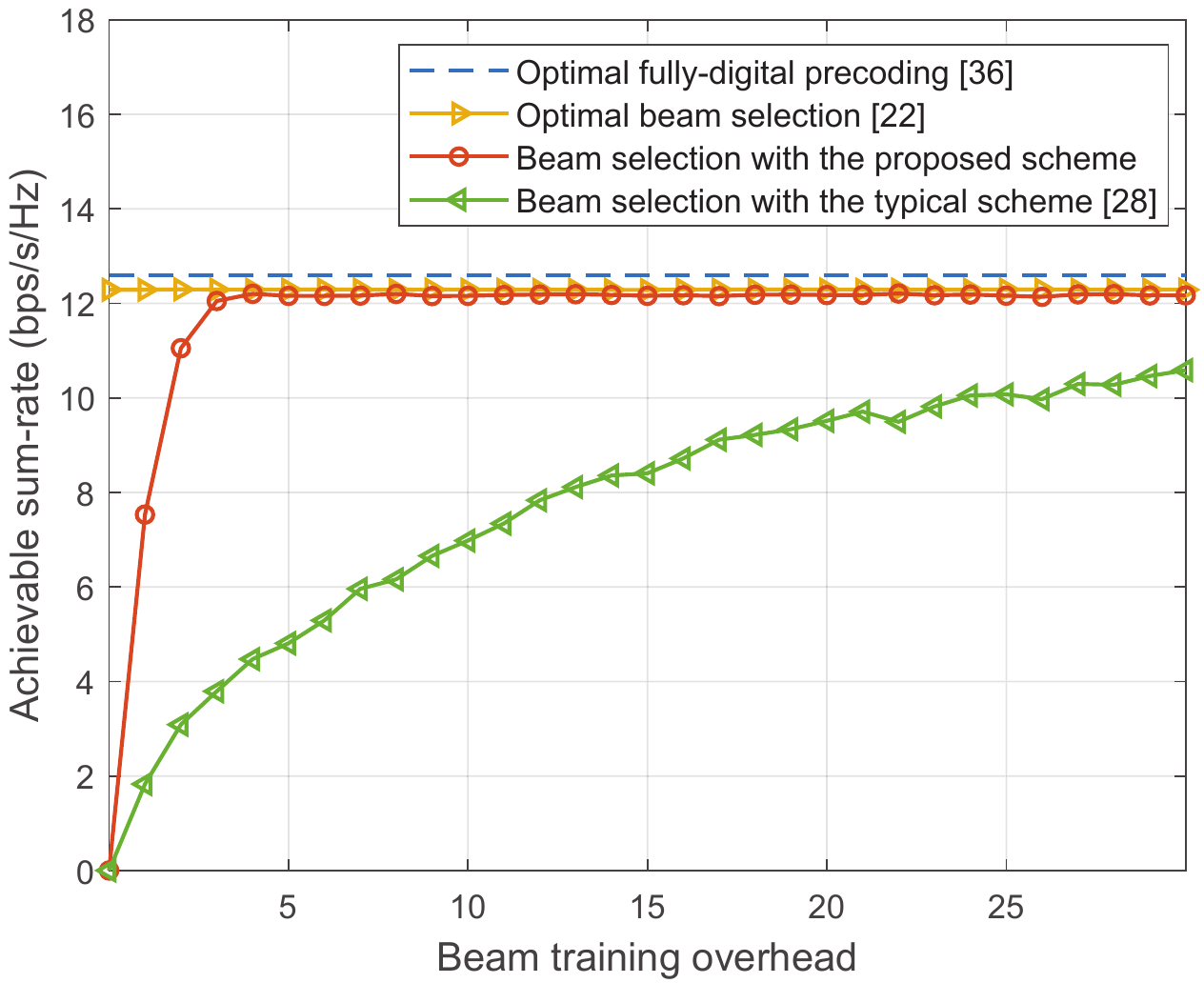}
	\caption{The achievable sum-rate performance against the beam training overhead $T$ with $f_\mathrm{c}=300$ GHz, $B=5$ GHz, and $\alpha_{k,\mathrm{max}}=0.2$.}
\end{figure}

Fig. 14 shows the achievable sum-rate performance against the beam training overhead $T$ with parameters $f_\mathrm{c}=300$ GHz, $B=5$ GHz and $\alpha_{k,\mathrm{max}}=0.2$. The other simulation settings are the same with that in Fig. 12 and Fig. 13. We can find similar observation in Fig. 14 compared with Fig. 12 and Fig. 13. Specifically, the proposed scheme can still achieve near-optimal achievable sum-rate performance, with a much smaller beam training overhead than the typical scheme. This indicates that when the beam split effect is relatively not serious with $f_\mathrm{c}=300$ GHz and $B=5$ GHz, the proposed scheme still has the ability to generate effective beams to cover the tracking range.
	
\begin{figure}
	\centering
	\includegraphics[width=0.46\textwidth]{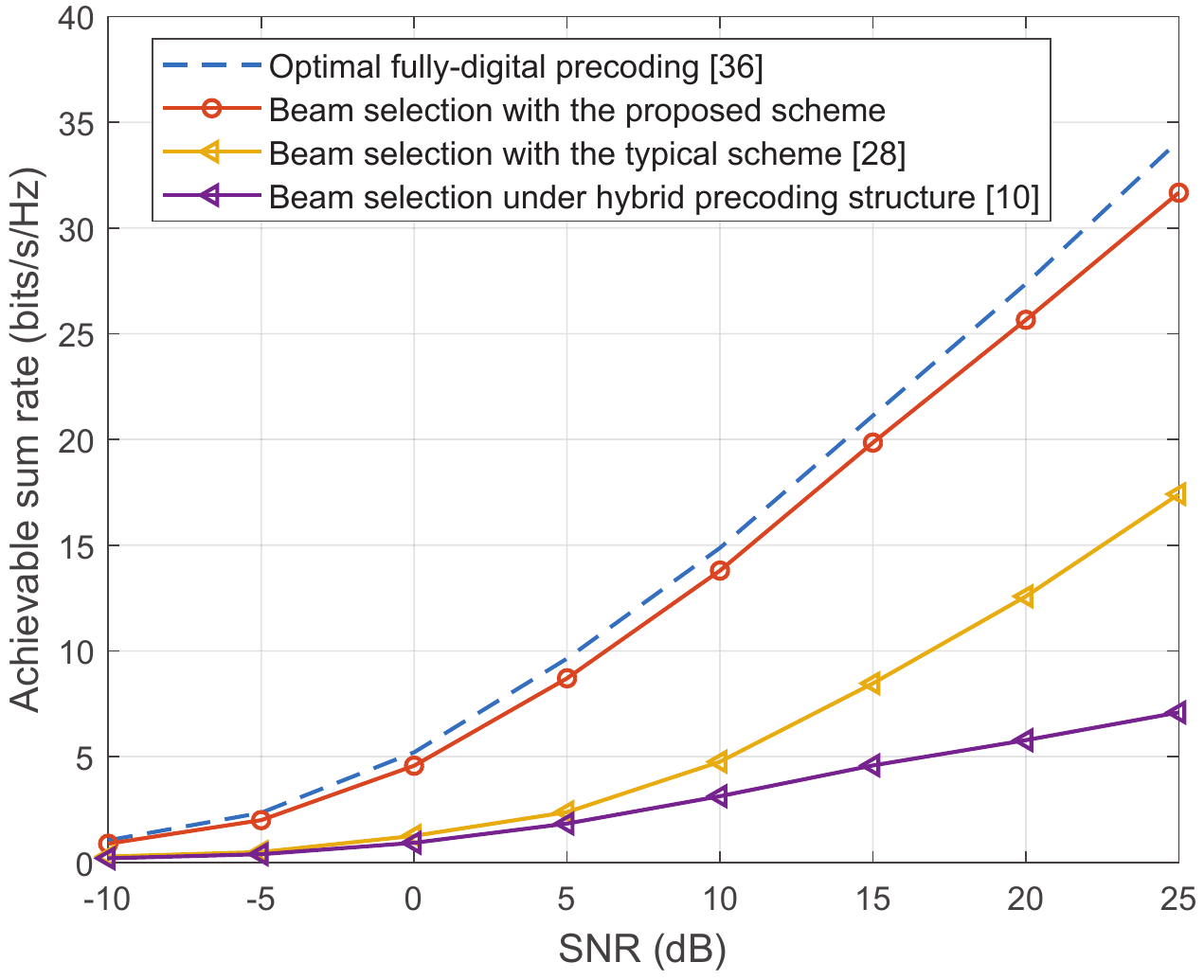}
	\caption{The achievable sum-rate performance comparison  against SNR.}                                                                                                                                                                                                                                                                                                                                                                                                                                                                                                                                                                                                                                                                                                                                                                                                                                                                                                                                                                                                                                                                                                                                                                                                                                                                                                                                                                                                                                                                                                                                                                                                                                                                                                                                                                                                                                                                                                                                                                                                                                                                   
\end{figure}	
	
Fig. 15 illustrates the achievable sum-rate performance against the SNR, where the performance of the beam selection based precoding method with the physical directions tracked by the proposed scheme and the beam selection based precoding method based on the physical directions tracked by the typical beam tracking scheme \cite{Ref:BeamAlignment2010} as described in Subsection \ref{Tra} are provided. Here, we set $T=5$ and $\alpha_{k,\mathrm{max}}=0.1$. The optimal fully-digital MMSE precoding \cite{Ref:MMSEPre2019} with the perfect channel information and the beam selection based precoding method with the perfect physical directions using hybrid precoding structure\cite{Ref:SpatiallyPre2014} are also depicted as benchmarks for comparison. We can observe from Fig. 15 that when $T=5$, the proposed beam zooming based beam tracking scheme outperforms the typical scheme, and it can obtain the near-optimal achievable sum-rate performance compared with the optimal fully-digital MMSE precoding. In addition, we can observe that the beam selection based precoding method based on the perfect physical directions suffers from a severe achievable sum-rate performance loss when using hybrid precoding structure. This is mainly caused by the severe array gain loss due to the beam split effect. The results in Fig. 15 also verify the effectiveness of the DPP structure in wideband THz massive MIMO systems. 
	
\begin{figure}
	\centering
		\includegraphics[width=0.45\textwidth]{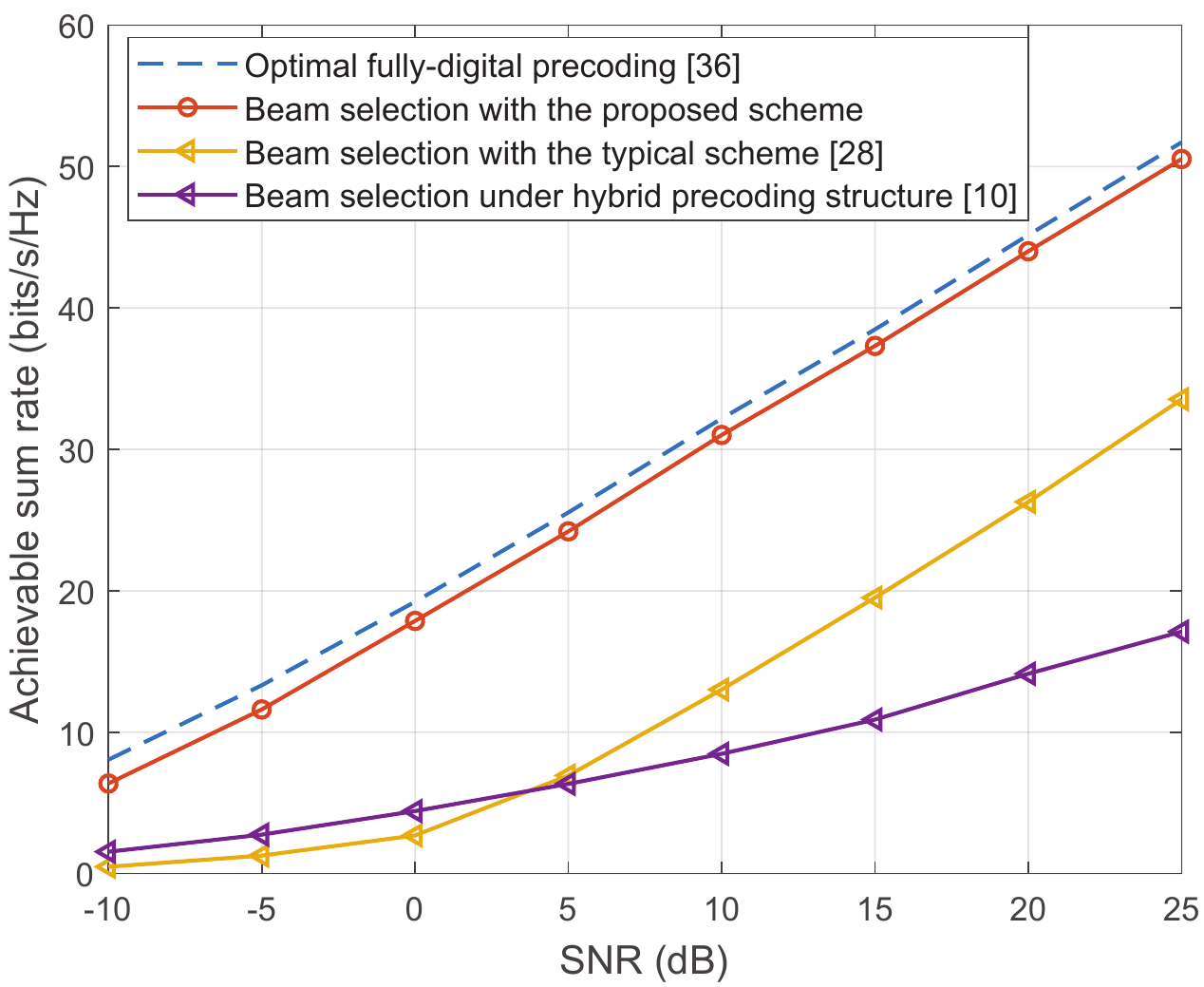}
		\caption{The achievable sum-rate performance comparison against SNR with multi-antenna users.}            
\end{figure}	
	
Fig. 16 illustrates the performance of the proposed beam zooming based beam tracking scheme in multi-antenna user case, where we consider the user parameters $N_\mathrm{r}=32$, $K_\mathrm{d}^\mathrm{r}=4$, $\alpha_{k}^\mathrm{r}=0.1$, and $T=8$ for beam tracking procedure at both the BS and the user side. The other system parameters are the same as that in Fig. 15. The optimal fully-digital precoding and beam selection under hybrid precoding scheme are carried out based on perfect channel information and perfect physical directions, respectively. We can observe from Fig. 16 that in multi-antenna user case, the proposed beam zooming based beam tracking scheme can still achieve near-optimal achievable sum-rate performance, which is also much better than that achieved by the typical scheme. Notice that the performance of beam selection with the typical scheme and beam selection under hybrid precoding structure have a crossing point when SNR = $5$ dB. This is because when the SNR is low, the beam tracking error of the typical beam tracking scheme becomes dominant, which causes more serious achievable sum-rate loss than that under hybrid precoding structure caused by the beam split effect. In contrast, when the SNR grows up, the typical beam tracking scheme can obtain correct physical directions with a high probability.  Under this circumstance, it can overcome the achievable rate loss caused by the beam split effect by utilizing the DPP structure, and thus realize better achievable sum-rate performance than that of beam selection under hybrid precoding structure.
	
By summarizing  simulation results in Fig. 9-16, we can conclude that the proposed beam zooming based beam tracking scheme can realize precise beam tracking with the low overhead and the near-optimal achievable sum-rate performance.

\section{Conclusions}\label{Con}
	
In this paper, we investigated the wideband beam tracking problem in THz massive MIMO systems. To realize accurate and fast beam tracking, at first we proved the beam zooming mechanism to flexibly control the angular coverage of beams generated by the DPP structure, i.e., the degree of the beam split effect. Based on the beam zooming mechanism, we then proposed a beam zooming based beam tracking scheme. Unlike traditional schemes where only one frequency-independent beam can be usually generated by one RF chain, the proposed scheme can simultaneously track multiple user physical directions by using multiple frequency-dependent beams generated by one RF chain, which is realized by flexibly controlling the degree of the beam split effect. Theoretical analysis and simulation results showed that the proposed scheme can track the user motion accurately with a reduced beam training overhead by about $90\%$ compared with the typical beam tracking scheme. Moreover, based on the physical directions tracked by the proposed scheme, nearly $98\%$ of the optimal sum-rate performance can be achieved, which makes the proposed beam zooming based beam tracking scheme attractive for THz massive MIMO systems. In the future, we will further investigate the beam tracking problem by considering some other THz channel characteristics, such as angle spread \cite{Ref:AngleSpread2017} and distance-aware effect\cite{Ref:THzDis2016}. Meanwhile, the beam tracking problem in reconfigurable intelligent surface aided THz massive MIMO systems also requires further investigation\cite{Ref:CERIS2021}.
	
\section*{Appendix A. Proof of Lemma 2}
\emph{Proof}: The beamforming vector $\mathbf{f}_{k,m}$ can be rewritten by substituting $\mathbf{A}_{k}^\mathrm{s}=\mathrm{blkdiag}\left(\mathbf{a}_{P}(\phi_{k})e^{j\pi(P\phi_{k}+2s_{k})\mathbf{p}^{T}(K_\mathrm{d})}\right)$ and $\mathbf{t}_{k}=s_{k}T_{c}\mathbf{p}(K_\mathrm{d})$ into $\mathbf{f}_{k,m}=\mathbf{A}_{k}^{s}e^{-j2\pi f_{m}\mathbf{t}_{k}}$ as
\begin{equation}\label{11-1}
\begin{aligned}
\mathbf{f}_{k,m}&=\mathrm{blkdiag}\left(\mathbf{a}_{P}(\phi_{k})e^{j\pi(P\phi_{k}+2s_{k})\mathbf{p}^{T}(K_\mathrm{d})}\right)e^{-j2\pi\xi_{m}s_{k}\mathbf{p}(K_\mathrm{d})}\\
&=\mathrm{blkdiag}\left(\mathbf{a}_{P}(\phi_{k})e^{j\pi P\phi_{k}\mathbf{p}^{T}(K_\mathrm{d})}\right)e^{j\pi(2s_{k}-2\xi_{m}s_{k})\mathbf{p}(K_\mathrm{d})}.
\end{aligned}
\end{equation}
We can observe that $\mathrm{blkdiag}\left(\mathbf{a}_{P}(\phi_{k})e^{j\pi P\phi_{k}\mathbf{p}^{T}(K_\mathrm{d})}\right)$ in (\ref{11-1}) has the same form as the analog beamformer $\mathbf{A}_{k}^\mathrm{s}$ in \textbf{Lemma 1}. Therefore, we can utilize \textbf{Lemma 1} to obtain the physical direction that the beam generated by  the beamforming vector $\mathbf{f}_{k,m}$ is aligned with. Specifically, if we make $e^{j\pi(2s_{k}-2\xi_{m}s_{k})\mathbf{p}(K_\mathrm{d})}$ in (\ref{11-1}) satisfy $e^{j\pi(2s_{k}-2\xi_{m}s_{k})\mathbf{p}(K_\mathrm{d})}=\left[1,e^{j\pi\beta_{k,m}},e^{j\pi 2\beta_{k,m}},\cdots,e^{j\pi(K_{d}-1)\beta_{k,m}}\right]^{T}$ in \textbf{Lemma 1}, the beamforming vector $\mathbf{f}_{k,m}$ has the same form as that in $\textbf{Lemma 1}$. Thus, the frequency-dependent phase shift $\beta_{k,m}$ can be represented by the number of periods that delayed by TDs $s_{k}$ as
\begin{equation}\label{12}
\beta_{k,m}=2(1-\xi_{m})s_{k}.
\end{equation}
Then, according to \textbf{Lemma 1}, the physical direction $\bar{\theta}_{k,m}$ that the beam generated by the beamforming vector $\mathbf{f}_{k,m}$ is aligned with, can be obtained by substituting $s_{k}$, $\phi_{k}$ and (\ref{12}) into (\ref{App1}) as
\begin{equation}\label{13}
\begin{aligned}
\bar{\theta}_{k,m}&=\frac{\phi_{k}}{\xi_{m}}-(\frac{1}{\xi_{m}}-1)(\phi_{k}+\frac{2\xi_{M}\xi_{1}\alpha_{k}}{\xi_{M}-\xi_{1}})\\
&=\theta_{k}^{(0)}+(1-\xi_{1})\alpha_{k}+\frac{2\xi_{M}\xi_{1}(\xi_{m}-1)}{\xi_{m}(\xi_{M}-\xi_{1})}\alpha_{k},
\end{aligned}
\end{equation}
which proves (\ref{10}).
	
Obviously, according to (\ref{13}), the physical direction $\bar{\theta}_{k,m}$ increases progressively as $m$ becomes larger. When $m=M$, we have
\begin{equation}\label{14}
\begin{aligned}
\bar{\theta}_{k,M}&=\theta_{k}^{(0)}+(1-\xi_{1})\alpha_{k}+\frac{2\xi_{M}\xi_{1}(\xi_{M}-1)}{\xi_{M}(\xi_{M}-\xi_{1})}\alpha_{k}\\
&\overset{(a)}{=}\theta_{k}^{(0)}+\alpha_{k},
\end{aligned}
\end{equation}
where $(a)$ comes from $\xi_{1}+\xi_{M}=\frac{f_{1}}{f_{c}}+\frac{f_{M}}{f_{c}}=2$. Similar to (\ref{14}), we can also obtain $\bar{\theta}_{k,1}=\theta_{k}^{(0)}-\alpha_{k}$ when $m=1$. Therefore, considering $\bar{\theta}_{k,m}$ is monotonously increasing over $m$, the physical directions $\bar{\theta}_{k,m},m=1,2,\cdots,M$ which the beams generated by beamforming vectors $\mathbf{f}_{k,m},m=1,2,\cdots,M$ are aligned with, can cover the whole angular tracking range $[\theta_{k}^{(0)}-\alpha_{k},\theta_{k}^{(0)}+\alpha_{k}]$.
$\hfill\blacksquare$
	
\section*{Appendix B. Lemma 3}
\begin{thm} \label{lemma3}
	When the DPP structure can eliminate the beam split effect at an arbitrary physical direction $\theta_{k,i}^{(0)}\in[-1,1]$, the following inequality holds
	\begin{equation}\label{AppB1}
	1\pm P\left(\xi_{m}-1\right)\left(\theta_{k,i}^{(0)}-\alpha_{k}\right)>0.
	\end{equation}	
\end{thm}
\emph{Proof:} Since the DPP structure should generate beams aligned with the physical direction $\theta_{k,i}^{(0)}$ during data transmission, it must have the ability to eliminate the beam split effect at an arbitrary physical direction $\theta_{k,i}^{(0)}\in[-1,1]$. That is, the limitation of the DPP structure, i.e., $\beta_{k,m}\in[-1,1]$, must hold for an arbitrary physical direction $\theta_{k,i}^{(0)}\in[-1,1]$. Considering that the frequency-dependent phase shifts $\beta_{k,m}$ satisfies $\beta_{k,m}\in[-1,1]$ in \textbf{Lemma 1}, and substituting $\beta_{k,m}=2(\xi_{m}-1)s_{k}$ in (\ref{12}) and $s_{k}=-\frac{P\theta_{k,i}^{(0)}}{2}$ into $\beta_{k,m}\in[-1,1]$, we have the following inequality that should be satisfied for an arbitrary physical direction $\theta_{k,i}^{(0)}$ as
\begin{equation}\label{AppB2}
-1\leq(\xi_{m}-1)\theta_{k,i}^{(0)} P\leq 1.
\end{equation}
Recalling that $\frac{f_{1}}{f_{c}}\leq\xi_{m}\leq\frac{f_{M}}{f_{c}}$ and $\theta_{k,i}^{(0)}\in[-1,1]$, we can obtain that the ratio between the number of antennas and the number of TDs $P=\frac{N}{K_\mathrm{d}}$ should satisfy
\begin{equation}\label{AppB3}
\left|(\xi_{m}-1)P\right|<1.
\end{equation}
From (\ref{15}), we know that $\theta_{k,i}^{(0)}-\alpha_{k}$ is the minimum element of the target physical direction set $\mathbf{\Psi}_{k}^{i+1}$, which obviously satisfies $|\theta_{k,i}^{(0)}-\alpha_{k}|\leq 1$. Therefore, (\ref{AppB3}) can be converted to
\begin{equation}\label{AppB4}
\left|P(\xi_{m}-1)(\theta_{k,i}^{(0)}-\alpha_{k})\right|<1,
\end{equation}
which is equivalent to (\ref{AppB1}).
$\hfill\blacksquare$

\bibliography{Ref}
\vspace{-15mm}
\begin{IEEEbiography}[{\includegraphics[width=1in,height=1.25in,clip,keepaspectratio]{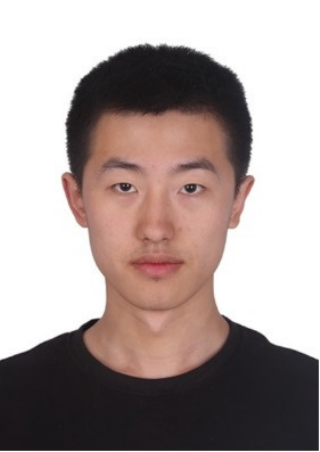}}]{Jingbo Tan} (Student Member, IEEE) received his B. S. degree in the Department of Electronic Engineering,  Tsinghua University, Beijing, China, in 2017, where he is currently pursuing his Ph. D. degree. His research interests include precoding and channel estimation in massive MIMO, THz communications, and reconfigurable intelligent surface aided systems. He has received the IEEE Communications Letters Exemplary Reviewer Award in 2018 and the Honorary Mention in the 2019 IEEE ComSoC Student Competition.
\end{IEEEbiography}
\vspace{-15mm}
\begin{IEEEbiography}[{\includegraphics[width=1in,height=1.25in,clip,keepaspectratio]{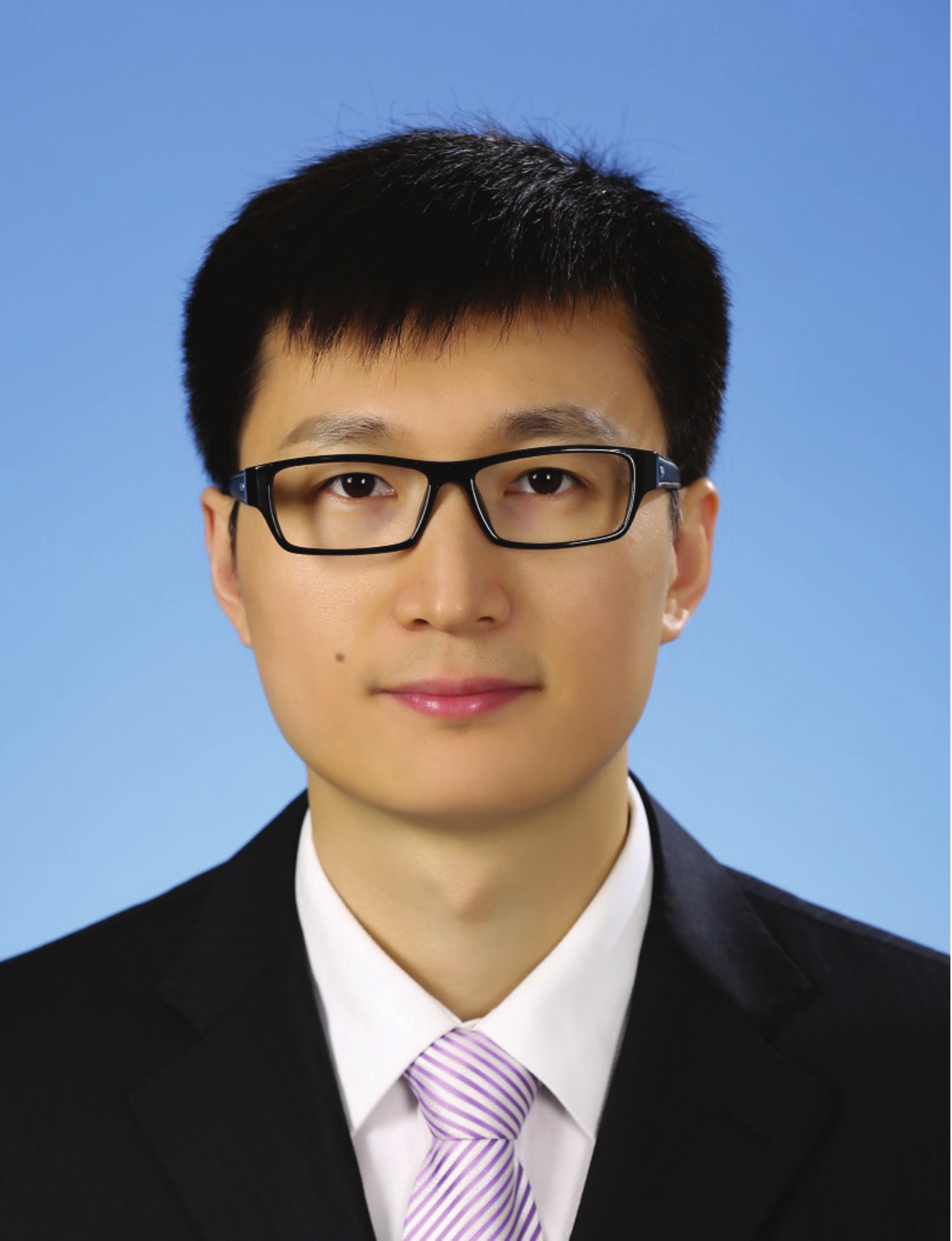}}]{Linglong Dai} (Senior Member, IEEE) received the B.S. degree from Zhejiang University, Hangzhou, China, in 2003, the M.S. degree (with the highest honor) from the China Academy of Telecommunications Technology, Beijing, China, in 2006, and the Ph.D. degree (with the highest honor) from Tsinghua University, Beijing, China, in 2011. From 2011 to 2013, he was a Postdoctoral Research Fellow with the Department of Electronic Engineering, Tsinghua University, where he was an Assistant Professor from 2013 to 2016 and has been an Associate Professor since 2016. His current research interests include massive MIMO, reconfigurable intelligent surface (RIS), millimeter-wave/Terahertz communications, and machine learning for wireless communications. He has coauthored the book “MmWave Massive MIMO: A Paradigm for 5G” (Academic Press, 2016). He has authored or coauthored over 60 IEEE journal papers and over 40 IEEE conference papers. He also holds 19 granted patents. He was listed as a Highly Cited Researcher by Clarivate in 2020. He has received five IEEE Best Paper Awards at the IEEE ICC 2013, the IEEE ICC 2014, the IEEE ICC 2017, the IEEE VTC 2017-Fall, and the IEEE ICC 2018. He has also received the Tsinghua University Outstanding Ph.D. Graduate Award in 2011, the Beijing Excellent Doctoral Dissertation Award in 2012, the China National Excellent Doctoral Dissertation Nomination Award in 2013, the URSI Young Scientist Award in 2014, the IEEE Transactions on Broadcasting Best Paper Award in 2015, the Electronics Letters Best Paper Award in 2016, the National Natural Science Foundation of China for Outstanding Young Scholars in 2017, the IEEE ComSoc Asia-Pacific Outstanding Young Researcher Award in 2017, the IEEE ComSoc Asia-Pacific Outstanding Paper Award in 2018, the China Communications Best Paper Award in 2019, and the IEEE Communications Society Leonard G. Abraham Prize in 2020. He is an Area Editor of IEEE Communications Letters, and an Editor of IEEE Transactions on Communications and IEEE Transactions on Vehicular Technology. Particularly, he is dedicated to reproducible research and has made a large amount of simulation code publicly available.
\end{IEEEbiography}

\end{document}